\documentclass[11pt,a4paper]{article}
\pdfoutput=1
\usepackage{jheppub}

\newcommand{\ah}{\hat a}
\newcommand{\cD}{{\cal D}}
\newcommand{\cO}{{\cal O}}
\newcommand{\as}{\alpha_s}
\newcommand{\wh}{\widehat}
\newcommand{\nn}{\nonumber}
\newcommand{\Cz}{C\!\!=\!0}

\newcommand{\ve}{\varepsilon}
\newcommand{\IM}{\mbox{\rm Im}}

\newcommand{\eqn}[1]{(\ref{#1})}

\newcommand{\svs}{\vbox{\vskip 4mm}}
\newcommand{\tvs}{\vbox{\vskip 6mm}}
\newcommand{\mvs}{\vbox{\vskip 8mm}}

\newcommand{\MSb}{{\overline{\rm MS}}}

\newcommand{\sfrac}[2]{\mbox{$\frac{#1}{#2}$}}

\title{Scalar correlator, Higgs decay into quarks, and scheme variations of
 the QCD coupling}

\author[a,b]{Matthias Jamin}
\author[a]{and Ramon Miravitllas}

\affiliation[a]{IFAE, BIST, Campus UAB, 08193 Bellaterra (Barcelona) Spain}

\affiliation[b]{ICREA, Pg. Llu\'\i s Companys 23, 08010 Barcelona, Spain}

\emailAdd{jamin@ifae.es}
\emailAdd{rmiravitllas@ifae.es}

\abstract{In this work, the perturbative QCD series of the scalar correlation
function $\Psi(s)$ is investigated. Besides $\IM\Psi(s)$, which is relevant for
Higgs decay into quarks, two other physical correlators, $\Psi^{''}(s)$ and
$D^L(s)$, have been employed in QCD applications like quark mass determinations
or hadronic $\tau$ decays. $D^L(s)$ suffers from large higher-order corrections
and, by resorting to the large-$\beta_0$ approximation, it is shown that this
is related to a spurious renormalon ambiguity at $u=1$. Hence, this correlator
should be avoided in phenomenological analyses. Moreover, it turns out
advantageous to express the quark mass factor, introduced to make the scalar
current renormalisation group invariant, in terms of the renormalisation
invariant quark mass $\wh m_q$. To further study the behaviour of the
perturbative expansion, we introduce a QCD coupling $\wh\alpha_s$, whose
running is explicitly renormalisation scheme independent. The scheme dependence
of $\wh\alpha_s$ is parametrised by a single parameter $C$, being related to
transformations of the QCD scale parameter $\Lambda$. It is demonstrated that
appropriate choices of $C$ lead to a substantial improvement in the behaviour
of the perturbative series for $\Psi^{''}(s)$ and $\IM\Psi(s)$.
}%


\begin{document}
\maketitle

\section{Introduction}\label{sect1}

The scalar correlation function in QCD plays an important role, as it governs
the decay of the Higgs into quark-antiquark pairs, and it has been employed
in determinations of quark masses from QCD sum rules as well as hadronic decays
of the $\tau$ lepton. Presently, the perturbative expansion for the scalar
correlator is known analytically up to order $\alpha_s^4$ in the strong
coupling \cite{bck05,che96,gkls90}, and estimates of the next, fifth order
have been attempted in the literature. While the decay of the Higgs boson
into quark-antiquark pairs is connected to the imaginary part of the scalar
correlator $\Psi(s)$ \cite{djou08}, two other physical correlators,
$\Psi^{''}(s)$ and $D^L(s)$, have been utilised in QCD sum rule analyses, the
former in quark mass extractions \cite{jop02,jop06} and the latter in hadronic
$\tau$ decays \cite{pp98,pp99,gjpps03}. In this work we shall investigate the
perturbative series of all three.

In order to achieve reliable error estimates of missing higher orders in QCD
predictions, a better understanding of the perturbative behaviour of the scalar
correlator at high orders is desirable. Work along those lines has been
performed in ref.~\cite{bkm00}, where the scalar correlation function has been
calculated in the large-$N_f$ approximation \cite{ben92,bro92}, or relatedly
the large-$\beta_0$ approximation \cite{bb94} (for a review see \cite{ben98}),
to all orders in the strong coupling.\footnote{For historical reasons, we shall
speak about the ``large-$\beta_0$'' approximation, although in the notation
employed in this work, the leading coefficient of the $\beta$-function is
termed $\beta_1$.} However, as will be discussed in more detail below, the
large-$\beta_0$ approximation does not provide a satisfactory representation
of the scalar correlator in full QCD. Still, as will be demonstrated, it can
serve as a guideline to shed light on the general structure of the scalar
correlation function.

Furthermore, while large QCD corrections are found in the case of the correlator
$D^L(s)$, the corrections are substantially smaller for $\IM\Psi(s)$ and
$\Psi^{''}(s)$. In the large-$\beta_0$ approximation this observation can
be traced back to the presence of a spurious renormalon pole in the Borel
transform at $u=1$ for $D^L(s)$, whereas $\Psi^{''}(s)$ and $\IM\Psi(s)$ are
free from this contribution. We discuss the origin of the additional renormalon
pole and its implications, but at any rate conclude that, in view of this fact,
the correlator $D^L(s)$ should be avoided in phenomenological analyses.

Additionally, the large-$\beta_0$ approximation motivates a strategy in order
to improve the perturbative expansion. The structure of the Borel transform in
the large-$\beta_0$ limit suggests the introduction of a renormalisation scheme
invariant QCD coupling $\wh\alpha_s$, which underlines the scheme invariance of
the perturbative term for the physical quantities under investigation. In fact,
all contributions of infrared (IR) and ultraviolet (UV) renormalons individually
are scheme independent. It is then found that higher-order corrections tend
to become smaller when re-expressing the perturbative series in terms of the
coupling $\wh\alpha_s$. One reason for this behaviour appears to be that
part of the perturbative corrections are resummed into a global prefactor
$\as^\delta$ which is present for the scalar correlator.

In full QCD, the construction of a scheme-invariant coupling does not appear
to be possible, at least in a universal sense, independent of any observable.
Nonetheless, we are able to provide the definition of a QCD coupling, which we
also term $\wh\alpha_s$, and whose running is scheme independent and described
by a simple $\beta$-function, only depending on the coefficients $\beta_1$ and
$\beta_2$. Different schemes can then be parametrised by a single parameter $C$,
which corresponds to transformations of the QCD scale parameter $\Lambda$. By
investigating two phenomenological applications, the correlator $\Psi^{''}(s)$
at the $\tau$ mass scale and $\IM\Psi(s)$ for Higgs decay to quarks, we show
that employing the coupling $\wh\alpha_s$ and choosing appropriate schemes by
varying the parameter $C$, the behaviour of the perturbative series can
be substantially improved.

Our article is organised as follows: in section~\ref{sect2}, theoretical
expressions for the scalar correlation function $\Psi(s)$ and the corresponding
physical correlators $\IM\Psi(s)$, $\Psi^{''}(s)$ and $D^L(s)$ are collected,
and the present knowledge on the perturbative expansions is summarised.
Furthermore, the renormalisation group invariant quark mass $\wh m_q$ is
introduced, and the correlators are rewritten in terms of this mass definition.
In section \ref{sect3}, we review the results of ref.~\cite{bkm00} on the
scalar correlation function in the large-$\beta_0$ approximation and apply
them to a discussion of the correlators $\Psi^{''}(s)$ and $D^L(s)$. Next, in
section \ref{sect4}, we define the coupling $\wh\alpha_s$, and compute its
$\beta$-function as well as the perturbative relation to $\alpha_s$ in the
$\MSb$ scheme. Finally, in section~\ref{sect5}, two phenomenological
applications, $\Psi^{''}(s)$ at the $\tau$ mass scale and $\IM\Psi(s)$ for
Higgs decay, are investigated, and followed by our conclusions in section
\ref{sect6}. More technical material like the coefficients of the
renormalisation group functions, higher-order coefficients relevant for the
large-$\beta_0$ approximation, as well as a discussion of the subtraction
constant $\Psi(0)$, are relegated to appendices.

\section{The scalar two-point correlator}\label{sect2}

The following work shall be concerned with the scalar two-point correlation
function $\Psi(p^2)$ which is defined by
\begin{equation}
\label{Psi}
\Psi(p^2) \,\equiv\, i\!\int\!{\rm d}x \,{\rm e}^{ipx} \langle\Omega|
T\{j(x) j^\dagger(0)\}|\Omega\rangle \,.
\end{equation}
The non-perturbative, full QCD vacuum is denoted by $|\Omega\rangle$.~For
our two applications, the scalar current $j(x)$ is chosen to arise either from
the divergence of the normal-ordered vector current,
\begin{equation}
\label{jtau}
j(x) \,=\, \partial^\mu \!:\!\bar u(x)\gamma_\mu s(x)\!: \;=\,
i\,(m_u-m_s) \!:\!\bar u(x) s(x)\!: \,,
\end{equation}
or the interaction of the Higgs boson with quarks,
\begin{equation}
j(x) \,=\, m_q \!:\!\bar q(x) q(x)\!: .
\end{equation}
These choices have the advantage of an additional factor of the quark masses,
which makes the currents $j(x)$ renormalisation group invariant (RGI).
Furthermore, the first current is taken to be flavour non-diagonal, with
a particular flavour content that plays a role in hadronic $\tau$ decays to
strange final states.\footnote{The $(\bar u d)$ flavour content that also
arises in hadronic $\tau$ decays is obtained by simply replacing the strange
with a down quark.} 

The purely perturbative expansion of $\Psi(p^2)$ is known up to order $\as^4$
\cite{bck05} and takes the general form
\begin{equation}
\label{PsiPT}
\Psi_{\rm PT}(s) \,=\, -\,\frac{N_c}{8\pi^2} \,m_\mu^2 \,s
\sum\limits_{n=0}^\infty a_\mu^n \sum\limits_{k=0}^{n+1} d_{n,k} L^k \,,
\end{equation}
where $s\equiv p^2$ and $a_\mu\equiv\as(\mu)/\pi$. To simplify the notation,
we have introduced the generic mass factor $m_\mu$ which either stands for the
combination $(m_u(\mu)-m_s(\mu))$ or $m_q(\mu)$.\footnote{In the case of a
flavour non-diagonal current, the so-called singlet-diagram contributions
are absent, and the perturbative expansion equally applies to the pseudoscalar
correlator, up to a replacement of the mass factor $(m_u-m_s)$ by $(m_u+m_s)$.}
The running quark masses and the QCD coupling are renormalised at the scale
$\mu$, which enters in $L\equiv\ln(-s/\mu^2)$. As a matter of principle,
different scales could be introduced for the renormalisation of coupling and
quark masses, but for simplicity, we refrain from this choice. Below, this
option will, however, be discussed in relation to renormalisation schemes.

At each perturbative order $n$, the only independent coefficients $d_{n,k}$
are the $d_{n,1}$. The coefficients $d_{n,0}$ depend on the renormalisation
prescription and do not contribute in physical quantities, while all remaining
coefficients $d_{n,k}$ with $k>1$ can be obtained by means of the
renormalisation group equation (RGE). The normalisation in eq.~\eqn{PsiPT} is
chosen such that $d_{0,1}=1$. Setting the number of colours $N_c=3$, and
employing the $\MSb$-scheme \cite{bbdm78}, after tremendous efforts the
coefficients $d_{n,1}$ up to $\cO(\as^4)$ were found to be
\cite{gkls90,che96,bck05}:
\begin{equation}
\label{d01tod21}
d_{0,1} \,=\, 1 \,, \qquad
d_{1,1} \,=\, \sfrac{17}{3} \,, \qquad
d_{2,1} \,=\, \sfrac{10801}{144} - \sfrac{39}{2} \zeta_3 +
\Big(\!- \sfrac{65}{24} + \sfrac{2}{3} \zeta_3 \Big) N_f
\end{equation}
\begin{displaymath}
\label{d31}
d_{3,1} \,=\, \sfrac{6163613}{5184} - \sfrac{109735}{216} \zeta_3 +
\sfrac{815}{12} \zeta_5 + \Big(\! -\sfrac{46147}{486} + \sfrac{262}{9} \zeta_3
- \sfrac{5}{6} \zeta_4 - \sfrac{25}{9} \zeta_5 \Big) N_f +
\Big( \sfrac{15511}{11664} - \sfrac{1}{3} \zeta_3 \Big) N_f^2 \nn
\end{displaymath}
\begin{eqnarray}
\label{d41}
d_{4,1} &\!=\!& \sfrac{10811054729}{497664} - \sfrac{3887351}{324} \zeta_3 +
\sfrac{458425}{432} \zeta_3^2 + \sfrac{265}{18} \zeta_4 +
\sfrac{373975}{432} \zeta_5 - \sfrac{1375}{32} \zeta_6 -
\sfrac{178045}{768} \zeta_7 \nn \\
\tvs
&& +\,\Big(\! - \sfrac{1045811915}{373248} + \sfrac{5747185}{5184} \zeta_3 -
\sfrac{955}{16} \zeta_3^2 - \sfrac{9131}{576} \zeta_4 +
\sfrac{41215}{432} \zeta_5 + \sfrac{2875}{288} \zeta_6 +
\sfrac{665}{72} \zeta_7 \Big) N_f \nn \\
\tvs
&& \hspace{-12mm}
+\,\Big( \sfrac{220313525}{2239488} - \sfrac{11875}{432} \zeta_3 +
\sfrac{5}{6} \zeta_3^2 + \sfrac{25}{96} \zeta_4 - \sfrac{5015}{432} \zeta_5
\Big) N_f^2 + \Big(\! -\sfrac{520771}{559872} + \sfrac{65}{432} \zeta_3 +
\sfrac{1}{144} \zeta_4 + \sfrac{5}{18} \zeta_5 \Big) N_f^3 \,. \nn
\end{eqnarray}
For future reference, at $N_f=3$, numerically, the respective coefficients take
the values
\begin{equation}
\label{d11tod41n}
d_{1,1} \,=\,  5.6667 \,, \qquad
d_{2,1} \,=\, 45.846  \,, \qquad
d_{3,1} \,=\, 465.85  \,, \qquad
d_{4,1} \,=\, 5588.7  \,.
\end{equation}
The case $N_f=5$, relevant for Higgs boson decay, will be considered in the
phenomenological applications of section~\ref{sect5}.

As indicated above, the correlator $\Psi(s)$ itself is not related to a
measurable quantity. Since it grows linearly with $s$ as $s$ tends to infinity,
it satisfies a dispersion relation with two subtraction constants,
\begin{equation}
\label{DisRelPsi}
\Psi(s) \,=\, \Psi(0) + s\,\Psi^{'}(0) + s^2
\!\int\limits_0^\infty \!\frac{\rho(s')}{(s')^2 (s'-s-i0)}\,{\rm d}s' \,,
\end{equation}
where $\rho(s)\equiv \IM\Psi(s+i0)/\pi$ is the scalar spectral function.
Hence, a possibility to construct a physical quantity other than the spectral
function itself, which will be discussed further down below, is to employ the
second derivative of $\Psi(s)$ with respect to $s$. Since the two derivatives
remove the two unphysical subtractions, $\Psi^{''}(s)$ is then only related to
the spectral function. The corresponding dispersion relation reads
\begin{equation}
\label{DisRelPsipp}
\Psi^{''}(s) \,=\, 2 \!\int\limits_0^\infty \!
\frac{\rho(s')}{(s'-s-i0)^3}\,{\rm d}s' \,,
\end{equation}
and the general perturbative expansion is
\begin{equation}
\label{PsippPT}
\Psi_{\rm PT}^{''}(s) \,=\, -\,\frac{N_c}{8\pi^2}\,\frac{m_\mu^2}{s}\,
\sum\limits_{n=0}^\infty a_\mu^n \sum\limits_{k=1}^{n+1} d_{n,k} \,k
\,\big[ L^{k-1} + (k-1) L^{k-2} \big] \,.
\end{equation}
Being a physical quantity, $\Psi^{''}(s)$ satisfies a homogeneous RGE, and
therefore the logarithms can be resummed with the particular scale choice
$\mu^2=-s\equiv Q^2$, leading to the compact expression
\begin{equation}
\label{Psippres}
\Psi_{\rm PT}^{''}(Q^2) \,=\, \frac{N_c}{8\pi^2}\,\frac{m_Q^2}{Q^2}\,\biggl\{
\, 1 + \sum\limits_{n=1}^\infty \,(d_{n,1} + 2 d_{n,2}) \,a_Q^n \,\biggr\} \,.
\end{equation}
In this way, both the running quark mass as well as the running QCD coupling
are to be evaluated at the renormalisation scale $Q$. The dependent
coefficients $d_{n,2}$ can be calculated from the RGE. They are collected in
appendix~\ref{appA}, together with the coefficients of the QCD $\beta$-function
and mass anomalous dimension. Numerically, at $N_f=3$, the perturbative
coefficients $d_{n,1}^{\,''}\equiv d_{n,1} + 2 d_{n,2}$ of eq.~\eqn{Psippres}
take the values
\begin{equation}
\label{dt11todt41n}
d_{1,1}^{\,''} \,=\,  3.6667 \,, \qquad
d_{2,1}^{\,''} \,=\, 14.179  \,, \qquad
d_{3,1}^{\,''} \,=\, 77.368  \,, \qquad
d_{4,1}^{\,''} \,=\, 511.83  \,.
\end{equation}
It is observed that the coefficients \eqn{dt11todt41n} for the physical
correlator are substantially smaller than the $d_{n,1}$ of eq.~\eqn{d11tod41n}.

For the ensuing discussion it will be advantageous to remove the running effects
of the quark mass from the remaining perturbative series. This can be achieved
by rewriting the running quark masses $m_q(\mu)$ in terms of RGI quark masses
$\wh m_q$ which are defined through the relation
\begin{equation}
\label{mhat}
m_q(\mu) \,\equiv\, \wh m_q \,[\alpha_s(\mu)]^{\gamma_m^{(1)}/\beta_1}
\exp\Biggl\{ \int\limits_0^{a_\mu} \!{\rm d}a \biggl[
\frac{\gamma_m(a)}{\beta(a)} - \frac{\gamma_m^{(1)}}{\beta_1 a} \biggr]
\Biggr\} \,.
\end{equation}
Accordingly, we define a modified perturbative expansion with new coefficients
$r_n$,
\begin{equation}
\label{Psippmhat}
\Psi_{\rm PT}^{''}(Q^2) \,=\, \frac{N_c}{8\pi^2}\,
\frac{\wh m^2}{Q^2} \,[\alpha_s(Q)]^{2\gamma_m^{(1)}/\beta_1}
\biggl\{\, 1 + \sum_{n=1}^\infty \, r_n \,a_Q^n \,\biggr\} \,,
\end{equation}
which now contain contributions from the exponential factor in eq.~\eqn{mhat}.
At $N_f=3$ the coefficients $r_n$ take the numerical values
\begin{equation}
\label{rnqcd}
r_1 \,=\, 5.4568 \,, \quad
r_2 \,=\, 24.287 \,, \quad
r_3 \,=\, 122.10 \,, \quad
r_4 \,=\, 748.09 \,.
\end{equation}
The order $\as^4$ coefficient $r_4$ depends on quark-mass anomalous dimensions
as well as $\beta$-function coefficients up to five-loops which for the
convenience of the reader in our conventions have been collected in
appendix~\ref{appA}.

As a second observable, we discuss the imaginary part of the scalar correlator
$\IM\Psi(s)$. After resumming the logarithms with the scale choice
$\mu^2=s\equiv M^2$, its general perturbative expansion reads
\begin{eqnarray}
\label{ImPsi}
\IM\Psi_{\rm PT}(s+i0) &=& \frac{N_c}{8\pi}\,m_M^2\, s\,
\sum_{n=0}^\infty a_M^n \sum_{l=0}^{[n/2]} d_{n,2l+1}\, (i\pi)^{2l} \\
\tvs
&=& \frac{N_c}{8\pi}\,m_M^2\, s\, \Big[\,1 + 5.6667\,a_M + 31.864\,a_M^2 +
89.156\,a_M^3 - 536.84\,a_M^4 + \ldots \Big] . \nn
\end{eqnarray}
In the first line, $[x]$ denotes the integer value of $x$, and in the second
line, the numerics has again been provided for $N_f=3$. We remark that in the
$\MSb$ scheme the fourth order coefficient turns out to be negative. However,
this does not necessarily imply an onset of the dominance of UV renormalons,
since the $(i\pi)^{2l}$ terms give a large contribution and contribute to the
sign change. Also for the imaginary part, we introduce a modified perturbative
series which results from rewriting the mass factor in terms of the invariant
quark mass. This yields
\begin{equation}
\label{ImPsihat}
\IM\Psi_{\rm PT}(s+i0) \,=\, \frac{N_c}{8\pi}\,\wh m^2\, s\,
[\alpha_s(M)]^{2\gamma_m^{(1)}/\beta_1}
\biggl\{\, 1 + \sum_{n=1}^\infty \, \bar r_n \,a_M^n \,\biggr\} \,.
\end{equation}
At $N_f=3$, this time the coefficients $\bar r_n$ assume the values
\begin{equation}
\label{rnbar}
\bar r_1 \,=\, 7.4568 \,, \quad
\bar r_2 \,=\, 45.552 \,, \quad
\bar r_3 \,=\, 172.64 \,, \quad
\bar r_4 \,=\, -\,204.09 \,.
\end{equation}

Besides $\Psi^{''}(s)$ and $\IM\Psi(s)$, in addition, below another physical
quantity shall be investigated, which is closer to the correlation functions
arising in hadronic $\tau$ decays. To this end, consider the general
decomposition of the vector correlation function into {\em transversal} ($T$)
and {\em longitudinal} ($L$) correlators:
\begin{eqnarray}
\label{PiVA}
\Pi_{\mu\nu}(p) &\equiv& i\!\int\!{\rm d}x \,{\rm e}^{ipx} \langle\Omega|
T\{j_\mu(x) j_\nu^\dagger(0)\}|\Omega\rangle  \,=\,
(p_\mu p_\nu - g_{\mu\nu} p^2)\, \Pi^T(p^2) + p_\mu p_\nu \,\Pi^L(p^2) \nn \\
\tvs
&=& (p_\mu p_\nu - g_{\mu\nu} p^2)\,\Pi^{T+L}(p^2) +
g_{\mu\nu} p^2\, \Pi^L(p^2) \,,
\end{eqnarray}
where $j_\mu(x)=\;:\!\!\bar u(x)\gamma_\mu s(x)\!\!:\,$. The correlators of
the decomposition in the second line, $\Pi^{T+L}(s)$ and $\Pi^L(s)$ are free
of kinematical singularities and thus should be employed in phenomenological
analyses. Next, the longitudinal correlator $\Pi^L(s)$ is related to the scalar
correlation function via
\begin{equation}
\label{PiL}
\Pi^L(s) \,=\, \frac{1}{s^2} \left[\, \Psi(s) - \Psi(0) \,\right] \,.
\end{equation}
Eq.~\eqn{PiL} suggests to define a third physical quantity
$D^L(s)$ by \cite{pp98,pp99,gjpps03}
\begin{equation}
\label{DL}
D^L(s) \,\equiv\, -\,s\,\frac{{\rm d}}{{\rm d} s} \Big[ s\,\Pi^L(s) \Big]
\,=\, \frac{1}{s} \left[\, \Psi(s) - \Psi(0) \,\right] - \Psi^\prime(s) \,.
\end{equation}

Employing eqs.~\eqn{PiL} and \eqn{DL}, together with the expansion \eqn{PsiPT},
the general form of the perturbative expansion for $D^L(Q^2)$ reads
\begin{equation}
\label{DLPT}
D_{\rm PT}^{L}(s) \,=\, -\,\frac{N_c}{8\pi^2}\, m_\mu^2\,
\sum\limits_{n=0}^\infty a_\mu^n \sum\limits_{k=1}^{n+1} k\,d_{n,k} L^{k-1}\,.
\end{equation}
Comparing eq.~\eqn{DLPT} to the corresponding expression for the Adler
function \cite{bj08}, one observes that up to the global prefactor --
which however depends on the scale dependent quark mass -- they are completely
equivalent. Being a physical quantity, also $D^L(s)$ satisfies a homogeneous
RGE, and thus again the logarithms in eq.~\eqn{DLPT} can be resummed with
the scale choice $\mu^2=-s=Q^2$, leading to the simple expression
\begin{equation}
\label{DLres}
D_{\rm PT}^{L}(Q^2) \,=\, -\,\frac{N_c}{8\pi^2} \,m_Q^2
\sum\limits_{n=0}^\infty d_{n,1} \,a_Q^n \,.
\end{equation}
From eq.~\eqn{DLres} it is again apparent that the only physically relevant
coefficients are the $d_{n,1}$. All the rest is encoded in running coupling
and quark masses. However, as only the $d_{n,1}$ enter, the perturbative
behaviour of $D^L(s)$ is substantially worse than that of the correlator
$\Psi^{''}(s)$. We shall shed further light on this observation in the next
section.

In analogy to eqs.~\eqn{Psippmhat} and \eqn{ImPsihat}, we can define a new
expansion by rewriting the running quark mass in terms of the RGI one. The
corresponding general perturbative expansion for $D^L(Q^2)$ reads
\begin{equation}
\label{DLmhat}
D_{\rm PT}^{L}(Q^2) \,=\, -\,\frac{N_c}{8\pi^2}\, \wh m^2\,
[\alpha_s(Q)]^{2\gamma_m^{(1)}/\beta_1} \biggl\{\, 1 +
\sum_{n=1}^\infty \,\tilde r_n a_Q^n \,\biggr\} \,,
\end{equation}
which defines the coefficients $\tilde r_n$. Numerically, at $N_f=3$, the
$\tilde r_n$ are found to be
\begin{equation}
\label{rntqcd}
\tilde r_1 \,=\, 7.4568 \,, \quad
\tilde r_2 \,=\, 59.534 \,, \quad
\tilde r_3 \,=\, 574.36 \,, \quad
\tilde r_4 \,=\, 6645.3 \,.
\end{equation}
As the next step, we review and utilise the information available on the scalar
correlation function in the large-$N_f$, or relatedly, the large-$\beta_0$
approximation.

\section{Large-\boldmath{$\beta_0$} approximation for the scalar correlator}\label{sect3}

The large-$\beta_0$ approximation for the scalar correlation function
was worked out in an impressive tour de force by Broadhurst et al.~in
ref.~\cite{bkm00}. The approach is to first calculate the large-$N_f$
expansion by summing fermion-loop chains in the gluon propagator, and then
performing the naive non-abelianisation \cite{bb94} through the replacement
$N_f \to -3\beta_1$. Taking into account that the correlator $\Pi_S(Q^2)$ of
\cite{bkm00} is related to $\Psi(Q^2)$ by $\Pi_S(Q^2)=(4\pi)^2 \Psi(Q^2)$,
in the large-$N_f$ limit the scalar correlator was found to be
\begin{equation}
\label{PsiLargeNf}
\Psi(Q^2) \,=\, \frac{N_c}{8\pi^2}\,m_\mu^2 \,Q^2 \biggl[\, L - 2 +
\frac{C_F b}{2T_F N_f}\,H(L,b) + \cO\biggl(\frac{1}{N_f^2}\biggr) +
\cO\biggl(\frac{1}{Q^2}\biggr) \,\biggr] \,.
\end{equation}
The function $H(L,b)$, with $b\equiv T_F N_f\,a_\mu/3$, is at the heart of the
work \cite{bkm00} and will be discussed in detail below.\footnote{Some care
has to be taken when implementing expressions from ref.~\cite{bkm00}, since
our convention for the logarithm is $L=\ln(Q^2/\mu^2)$, while in \cite{bkm00}
instead $\ln(\mu^2/Q^2)$ was employed.} In our conventions, $T_F=1/2$.

Comparing eqs.~\eqn{PsiPT} and \eqn{PsiLargeNf}, it immediately follows that
\begin{equation}
\sum\limits_{n=1}^\infty a_\mu^n \sum\limits_{k=0}^{n+1} d_{n,k} L^k \,=\,
\frac{C_F b}{2T_F N_f}\,H(L,b) \,.
\end{equation}
Next, employing the expansion
\begin{equation}
H(L,b) \,=\, \sum_{n=1}^\infty H_{n+1}(L)\, b^{n-1},
\end{equation}
along the lines of ref.~\cite{bkm00}, one obtains
\begin{equation}
\sum\limits_{k=0}^{n+1} d_{n,k} L^k \,=\, C_F\,\frac{N_f^{n-1}}{6^n}\,
H_{n+1}(L) \,,
\end{equation}
and in particular
\begin{equation}
\label{dn1LargeNf}
d_{n,1} \,=\, C_F\,\frac{N_f^{n-1}}{6^n}\, H_{n+1}^{(1)} \,,
\end{equation}
for the independent coefficients $d_{n,1}$, where $H_{n+1}^{(1)}$ denotes the
coefficient of the term of $H_{n+1}(L)$ linear in the logarithm. It remains
to arrive at an expression for $H_{n+1}^{(1)}$.

An explicit expression for the $H_{n+1}^{(1)}$ can be pieced together from
several formulae presented in ref.~\cite{bkm00}, the central of which, for
$n \geq 1$, reads:
\begin{equation}
\label{Hnp1L}
n(n+1) H_{n+1}(L) \,=\, (n+1)\big[ h_{n+2} + 4 (L-2) g_{n+1} \big] +
4 g_{n+2} +  9\,(-1)^n\,\cD_{n+1}(L) \,.
\end{equation}
The coefficients $h_{n+2}$ are scheme-dependent constants, which do not concern
us here since they are independent of $L$, while the quantities $g_n$ are
related to the expansion coefficients of the quark-mass anomalous dimension
$\gamma_m(a)$ in the large-$N_f$ limit. In this limit, one finds
\cite{gra93,bkm00}
\begin{equation}
\label{gammam}
\gamma_m(a) \,\equiv\, -\,\frac{\mu}{m_\mu}\frac{{\rm d}m_\mu}{{\rm d}\mu}\,=\,
\frac{2\,C_F b}{T_F N_f}\,g(b) + \cO\biggl(\frac{1}{N_f^2}\biggr) \,,
\end{equation}
with the function $g(b)$ being given by
\begin{equation}
\label{gb}
g(b) \,=\, \frac{(3-2b)^2}{(4-2b)}\,\frac{\Gamma(2-2b)}{[\Gamma(2-b)]^2}\,
\frac{\sin(\pi b)}{\pi b} \,.
\end{equation}
Then, finally, the expansion of $g(b)$, together with an efficient way to
generate it, which was also presented in \cite{bkm00}, reads:
\begin{equation}
\label{gbn}
g(b) \,=\, \sum\limits_{n=1}^\infty g_n b^{n-1} \,=\,
\Biggl[ 4 - \sum\limits_{n=2}^\infty\biggl(\frac{3}{2^n}+\frac{n}{2}\biggr)
b^{n-2} \Biggr] \exp\Biggl( \sum\limits_{l=3}^\infty \frac{2^l-3-(-1)^l}{l}\,
\zeta_l \,b^l \Biggr) \,.
\end{equation}
For the convenience of the reader, we list the first six coefficients $g_n$:
\begin{eqnarray}
\label{gnnum}
g_1  &=&  \sfrac{9}{4} \,, \qquad
g_2 \,=\, -\,\sfrac{15}{8} \,, \qquad
g_3 \,=\, -\,\sfrac{35}{16} \,, \qquad
g_4 \,=\, -\,\sfrac{83}{32} + \sfrac{9}{2}\,\zeta_3 \,, \nn \\
\tvs
g_5  &=&  -\,\sfrac{195}{64} - \sfrac{15}{4}\,\zeta_3 + \sfrac{27}{4}\,\zeta_4
\,, \qquad
g_6 \,=\, -\,\sfrac{451}{128} - \sfrac{35}{8}\,\zeta_3 - \sfrac{45}{8}\,\zeta_4
+ \sfrac{27}{2}\,\zeta_5 \,.
\end{eqnarray}
Comparing the general expansion of $g(b)$ with the one for $\gamma_m(a)$,
the relation for the individual expansion coefficients is given by
\begin{equation}
\gamma_m^{(n)} \,=\, 4\,C_F\,\frac{N_f^{n-1}}{6^n}\, g_n \,.
\end{equation}
Employing the coefficients $g_n$ of eq.~\eqn{gnnum}, it can easily be verified
that the terms with the highest power in $N_f$ of $\gamma_m^{(n)}$ in
eq.~\eqn{gfun} are indeed reproduced.

The functions $\cD_n(L)$ in the last summand of \eqn{Hnp1L}, and the
corresponding coefficients $\cD_n^{(1)}$ linear in $L$, can be derived from
the following relation:\footnote{The relation to the corresponding coefficients
$\tilde\Delta_n$ of \cite{bkm00} is given by
$n(n-1)\tilde\Delta_n = -\,2\,\cD_n^{(1)}$.}
\begin{equation}
\label{Dn1L}
\sum\limits_{n=0}^\infty \frac{\cD_n(L)}{n!}\,u^n \,=\,
\big[\, 1 + u\,G_D(u) \big] \,{\rm e}^{-(L-5/3)u} \,.
\end{equation}
The term ``$-5/3$'' in the exponent is particular for the $\MSb$ scheme which
is employed unless otherwise stated. Below, we shall, however, generalise our
expressions to an arbitrary scheme for the coupling.  Furthermore, the function
$G_D(u)$ was found to be \cite{bkm00}
\begin{eqnarray}
G_D(u) &=& \frac{2}{1-u} - \frac{1}{2-u} +
\frac{2}{3} \sum\limits_{p=3}^\infty \frac{(-1)^p}{(p-u)^2} -
\frac{2}{3} \sum\limits_{p=1}^\infty \frac{(-1)^p}{(p+u)^2} \nn \\
\mvs
&=& \frac{2}{1-u} - \frac{1}{2-u} + \frac{1}{6} \Big[\,
\zeta\big(2,2-\sfrac{u}{2}\big) - \zeta\big(2,\sfrac{3}{2}-\sfrac{u}{2}\big) -
\zeta\big(2,1+\sfrac{u}{2}\big) + \zeta\big(2,\sfrac{1}{2}+\sfrac{u}{2}\big)
\,\Big] \nn \\
\mvs
\label{GDu}
&=& \sum\limits_{k>0} \frac{k+3}{3}\,(2-2^{-k})\,u^{k-1} -
\frac{8}{3} \sum\limits_{l>0} \zeta_{2l+1} l(1-4^{-l})\,u^{2l-1} \,.
\end{eqnarray}
The first line of eq.~\eqn{GDu} explicitly displays the renormalon structure,
separated in IR renormalon poles at positive integer $u$, and UV renormalon
poles at negative integer $u$, while the second gives an expression in terms
of the Hurwitz $\zeta$-function. Finally, the third line provides the Taylor
expansion of $G_D(u)$ around $u=0$, which corresponds to the perturbative
expansion.  Inserting the extracted coefficients $g_n$ and $\cD_n^{(1)}$ into
$H_{n+1}^{(1)}$ derived from eq.~\eqn{Hnp1L}, it is a simple matter to verify
that eq.~\eqn{dn1LargeNf} reproduces the contributions with the highest power
of $N_f$ in the coefficients $d_{n,1}$ of \eqn{d01tod21} for $n\geq 1$. To
facilitate the comparison, the first few coefficients $\cD_n^{(1)}$ and
$H_{n+1}^{(1)}$ have been collected in appendix~\ref{appB}.

Next, an expression for $\Psi^{''}(Q^2)$ of eq.~\eqn{PsippPT} in the
large-$\beta_0$ limit shall be derived. The required second derivative of
the function $H(L,b)$ with respect to $L$ can be extracted from expressions
provided in ref.~\cite{bkm00}, along the lines of the computation above which
led to the coefficients $d_{n,1}$. To convert the large-$N_f$ expansion into
the large-$\beta_0$ (or large-$\beta_1$) limit, all occurrences of $N_f$ have
to be replaced by $-3\beta_1$. Finally, rewriting sums over the $\cD_n$
coefficients (and derivatives) in terms of the Borel transform of the coupling,
those sums can be expressed in closed form containing the function $G_D(u)$.
This yields
\begin{equation}
\begin{aligned}
\label{Psipplb0}
\Psi_{\beta_0}^{''}(Q^2) \,=\, &\frac{N_c}{8\pi^2}\,
\frac{m_\mu^2}{Q^2} \biggl\{\, 1 - \frac{2}{\beta_1} \sum_{n=1}^\infty
\frac{\gamma_m^{(n+1)}}{n}\, a_\mu^n \,+ \\
\mvs
&\frac{3C_F}{\beta_1}\! \int\limits_0^\infty \!{\rm d}u\,
{\rm e}^{-2u/(\beta_1 a_\mu)} \Big[ (1-u)\big[ 1+u\,G_D(u) \big]
{\rm e}^{-(L-5/3)u} - 1 \Big] \frac{1}{u} + \ldots \,\biggr\} ,
\end{aligned}
\end{equation}
where the ellipses stand for terms with additional suppression in $\beta_1$
or $Q^2$. Because the integrand contains IR renormalon poles along the path of
integration, a prescription has to be specified in order to define the integral.
In the present study the principal-value prescription shall always be adopted.

As $\Psi^{''}(Q^2)$ satisfies a homogeneous RGE, the logarithm can be resummed
through the scale choice $\mu^2=Q^2$. Furthermore, the running of the quark
mass is reflected in the terms containing the coefficients of the quark-mass
anomalous dimension $\gamma_m^{(n)}$, except for the leading-order running
$\gamma_m^{(1)}$ which is cancelled by the last term ``$-1$'' in the square
brackets. Hence, the mass running (except for the leading order) can be resummed
by expressing the quark mass in terms of the RGI quark mass $\wh m$ according
to eq.~\eqn{mhat}. In addition, we rewrite the expression in terms of a
coupling $a_Q^C\equiv \alpha_s^C(Q)/\pi$ parametrised by a constant $C$,
specifying the renormalisation scheme and being defined by the relation:
\begin{equation}
\label{ahatlb0}
\frac{1}{\hat a_Q} \,\equiv\, \frac{1}{a_Q^C} + C\,\frac{\beta_1}{2}
\,=\, \frac{1}{a_Q^{\MSb}} - \frac{5}{3}\,\frac{\beta_1}{2} \,.
\end{equation}
The coupling $\hat a_Q$ for $C=0$ can be considered a scheme-independent
coupling at large-$\beta_0$. This leads to our final formula for
$\Psi^{''}(Q^2)$ in the large-$\beta_0$ approximation:
\begin{eqnarray}
\label{Psipplb0res}
\Psi_{\beta_0}^{''}(Q^2) &=& \frac{N_c}{8\pi^2}\,
\frac{\wh m^2}{Q^2} \,[\alpha_s^{C_m}(Q)]^{2\gamma_m^{(1)}/\beta_1} \biggl\{\,
1 - 2\,\frac{\gamma_m^{(1)}}{\beta_1} \ln\Big[ 1 + C_m \sfrac{\beta_1}{2}\,
a_Q^{C_m} \Big] \nn \\
\mvs
&& +\,\frac{2\pi}{\beta_1}\! \int\limits_0^\infty \!{\rm d}u\,
{\rm e}^{-2u/(\beta_1 a_Q^{C_a})} B[\Psi^{''}](u) + \ldots \,\biggr\} \,, 
\end{eqnarray}
where we have introduced two separate constants $C_m$ and $C_a$, referring to
the scheme dependencies of quark mass and coupling, respectively. The Borel
transform $B[\Psi^{''}](u)$ is given by
\begin{eqnarray}
\label{BPsipp}
B[\Psi^{''}](u) &=& \frac{3C_F}{2\pi}\,{\rm e}^{-C_a u}\,\Big[\,
(1-u)\,G_D(u) - 1 \,\Big] \nn \\
\mvs
&=& \frac{3C_F}{2\pi}\,{\rm e}^{-C_a u}\, \biggl\{\,
\frac{1}{(2-u)} - \frac{2}{3}\,\sum\limits_{p=3}^\infty\, (-1)^p
\biggl[\, \frac{(p-1)}{(p-u)^2} - \frac{1}{(p-u)} \,\biggr] \nn \\
&& \hspace{36.3mm} -\,\frac{2}{3}\,\sum\limits_{p=1}^\infty\, (-1)^p
\biggl[\, \frac{(p+1)}{(p+u)^2} - \frac{1}{(p+u)} \,\biggr] \,\biggr\} \,.
\end{eqnarray}
The second equality again provides the separation of the Borel transform
$B[\Psi^{''}](u)$ in IR and UV renormalon poles. The found general structure
is analogous to the one of the Adler function \cite{ben98}. Except for the
linear IR pole at $u=2$, being related to the gluon condensate, we have
quadratic and linear IR poles at all integer $u\geq 3$. Furthermore, quadratic
and linear UV renormalon poles are found for all integer $u\leq -1$. Hence,
like for the Adler function, at large orders the perturbative coefficients will
be dominated by the quadratic UV renormalon pole at $u=-1$ which lies closest
to $u=0$. As is also observed from eq.~\eqn{Psipplb0res}, the perturbative
series contains a term without renormalon singularities which is related to the
scheme dependence of the global prefactor $\alpha_s^C(Q)$. This ``no-pole''
contribution is absent in the scheme with $C=0$, in which the prefactor is
expressed in terms of the invariant coupling $\hat\alpha_s(Q)$.

Let us proceed to an investigation of the perturbative expansion for three
different choices of the renormalisation scheme. We begin with the $\MSb$
scheme for both mass and coupling, in which $C_m=C_a=-5/3$, and the coefficients
$r_n^{\beta_0}$, introduced in eq.~\eqn{Psippmhat}, are found to be
\begin{eqnarray}
\label{rnlb0MSb}
r_1^{\beta_0}(\MSb,\MSb) &=& \frac{16}{3} \,=\, 5.3333 \,, \quad
r_2^{\beta_0}(\MSb,\MSb) \,=\, \Big(\,\frac{143}{36} - 2\zeta_3\Big)
\beta_1 \,=\, 7.0565 \,, \nn \\
\mvs
r_3^{\beta_0}(\MSb,\MSb) &=& \Big(\,\frac{1465}{324} -
\frac{4}{3}\zeta_3 \Big) \beta_1^2 \,=\, 59.107 \,, \\
\mvs
r_4^{\beta_0}(\MSb,\MSb) \!&=&\! \Big(\,\frac{17597}{2592} +
\frac{5}{6}\zeta_3 - \frac{15}{2}\zeta_5 \Big) \beta_1^3 \,=\, 1.2504 \,. \nn
\end{eqnarray}
The first entry in the argument of $r_n^{\beta_0}$ refers to the scheme for the
mass and the second for the coupling. The numerical values have been given for
$N_f=3$. Comparing to eq.~\eqn{rnqcd}, except for the first coefficient $r_1$,
the higher-order coefficients are not at all well represented by the
large-$\beta_0$ approximation, with a complete failure observed at the fourth
order. To obtain a better understanding of this behaviour, the contribution
of the lowest-lying renormalon poles to the perturbative large-$\beta_0$
coefficients shall be investigated.

\begin{table}[t]
\begin{center}
\begin{tabular}{lrrrrrr}
\hline\hline
& $\quad r_1^{\beta_0}\;$ & $\quad r_2^{\beta_0}\;$ &
  $\quad r_3^{\beta_0}\;$ & $\quad r_4^{\beta_0}\;$ &
  $\quad r_5^{\beta_0}\;$ & $\quad r_6^{\beta_0}\;$ \\
\hline
${\rm UV}_{-1}$ & 25.0 & -56.7 & 31.7 & -15025.0 & 65.2 & 133.9 \\
${\rm UV}_{-2}$ & -6.2 & 3.5 & 1.1 & 618.5 & 0.4 & -0.8 \\
${\rm IR}_{2}$ & 18.8 & 69.1 & 42.3 & 10973.5 & 28.4 & -28.4 \\
${\rm IR}_{3}$ & -2.8 & -6.3 & -1.3 & 349.9 & 2.5 & -3.8 \\
${\rm IR}_{4}$ & 1.6 & 3.1 & 0.3 & -259.0 & -1.3 & 1.7 \\
No-Pole & 62.5 & 88.6 & 26.4 & 3514.5 & 4.6 & -2.2 \\
\hline
SUM & 98.8 & 101.3 & 100.7 & 172.4 & 99.8 & 100.4 \\
\hline\hline
& $\quad r_7^{\beta_0}\;$ & $\quad r_8^{\beta_0}\;$ &
  $\quad r_9^{\beta_0}\;$ & $\quad r_{10}^{\beta_0}\;$ &
  $\quad r_{11}^{\beta_0}\;$ & $\quad r_{12}^{\beta_0}\;$ \\
\hline
${\rm UV}_{-1}$ & 89.6 & 105.7 & 97.7 & 101.1 & 99.5 & 100.2 \\
${\rm UV}_{-2}$ & 0.0 & -0.1 & 0.0 & 0.0 & 0.0 & 0.0 \\
${\rm IR}_{2}$ & 9.0 & -4.9 & 2.1 & -1.0 & 0.4 & -0.2 \\
${\rm IR}_{3}$ & 1.5 & -0.8 & 0.3 & -0.1 & 0.1 & 0.0 \\
${\rm IR}_{4}$ & -0.6 & 0.3 & -0.1 & 0.0 & 0.0 & 0.0 \\
No-Pole & 0.3 & -0.1 & 0.0 & 0.0 & 0.0 & 0.0 \\
\hline
SUM & 99.9 & 100.1 & 100.0 & 100.0 & 100.0 & 100.0 \\
\hline\hline
\end{tabular}
\caption{Contribution (in percent) of the lowest-lying ultraviolet (UV) and
infrared (IR) renormalon poles as well as the no-pole term to the first 12
perturbative coefficients $r_n^{\beta_0}$ in the $\MSb$ scheme for both quark
mass and renormalon terms.\label{tab1}}
\end{center}
\end{table}

In table~\ref{tab1}, the contributions in percent of the two lowest-lying UV
renormalon poles at $u=-1,-2$ and three lowest-lying IR renormalon poles at
$u=2,3,4$ as well as the no-pole term to the first 12 perturbative coefficients
$r_n$ in the large-$\beta_0$ approximation and the $\MSb$ scheme are presented.
It is observed that starting with about the 5th order, the dominance of the
lowest-lying UV pole at $u=-1$ sets in. For the first two orders, the no-pole
term which does not contain a renormalon singularity, dominates. Furthermore,
for the 4th order, huge cancellations between the different contributions take
place. At this order, only when adding the no-pole term and UV and IR
renormalon contributions up to order $p=15$, a 1\% precision on the coefficient
$r_4^{\beta_0}$ is reached.

\begin{table}[t]
\begin{center}
\begin{tabular}{lrrrrrr}
\hline\hline
& $\quad r_1^{\beta_0}\;$ & $\quad r_2^{\beta_0}\;$ &
  $\quad r_3^{\beta_0}\;$ & $\quad r_4^{\beta_0}\;$ &
  $\quad r_5^{\beta_0}\;$ & $\quad r_6^{\beta_0}\;$ \\
\hline
${\rm UV}_{-1}$ & 66.7 & -496.0 & 43.1 & 440.0 & 68.4 & 131.0 \\
${\rm UV}_{-2}$ & -16.7 & 31.0 & 1.5 & -18.1 & 0.5 & -0.8 \\
${\rm IR}_{2}$ & 50.0 & 604.5 & 57.6 & -321.4 & 29.7 & -27.8 \\
${\rm IR}_{3}$ & -7.4 & -55.1 & -1.7 & -10.2 & 2.6 & -3.7 \\
${\rm IR}_{4}$ & 4.2 & 27.1 & 0.5 & 7.6 & -1.4 & 1.7 \\
\hline
SUM & 96.8 & 111.5 & 100.9 & 97.9 & 99.8 & 100.4 \\
\hline\hline
& $\quad r_7^{\beta_0}\;$ & $\quad r_8^{\beta_0}\;$ &
  $\quad r_9^{\beta_0}\;$ & $\quad r_{10}^{\beta_0}\;$ &
  $\quad r_{11}^{\beta_0}\;$ & $\quad r_{12}^{\beta_0}\;$ \\
\hline
${\rm UV}_{-1}$ & 89.9 & 105.6 & 97.7 & 101.1 & 99.5 & 100.2 \\
${\rm UV}_{-2}$ & 0.0 & -0.1 & 0.0 & 0.0 & 0.0 & 0.0 \\
${\rm IR}_{2}$ & 9.1 & -4.9 & 2.1 & -1.0 & 0.4 & -0.2 \\
${\rm IR}_{3}$ & 1.5 & -0.8 & 0.3 & -0.1 & 0.1 & 0.0 \\
${\rm IR}_{4}$ & -0.6 & 0.3 & -0.1 & 0.0 & 0.0 & 0.0 \\
\hline
SUM & 99.9 & 100.1 & 100.0 & 100.0 & 100.0 & 100.0 \\
\hline\hline
\end{tabular}
\caption{Contribution (in percent) of the lowest-lying ultraviolet (UV) and
infrared (IR) renormalon poles to the first 12 perturbative coefficients
$r_n^{\beta_0}$ in the mixed scheme with $C_m=0$ for the quark mass and
$\MSb$ in the renormalon terms.\label{tab2}}
\end{center}
\end{table}

Now, we move to the discussion of renormalisation schemes for which the mass
renormalisation is taken at $C_m=0$, and thus the no-pole, logarithmic term of
eq.~\eqn{Psipplb0res} vanishes. Since the renormalisation scheme in the mass
and in the renormalon contribution can be chosen independently, we still have
the freedom to employ a different scheme in the latter case. Using the $\MSb$
scheme in the Borel integral, $C_a=-5/3$, the first four perturbative
coefficients are found to be:
\begin{eqnarray}
\label{rnlb0Mix}
r_1^{\beta_0}(\Cz,\MSb) &=& 2 \,, \quad
r_2^{\beta_0}(\Cz,\MSb) \,=\, \Big(\,\frac{31}{12} - 2\zeta_3\Big)
\beta_1 \,=\, 0.8065 \,, \nn \\
\mvs
r_3^{\beta_0}(\Cz,\MSb) &=& \Big(\,\frac{15}{4} -
\frac{4}{3}\zeta_3 \Big) \beta_1^2 \,=\, 43.482 \,, \\
\mvs
r_4^{\beta_0}(\Cz,\MSb) \!&=&\! \Big(\,\frac{5449}{864} +
\frac{5}{6}\zeta_3 - \frac{15}{2}\zeta_5 \Big) \beta_1^3 \,=\, -\,42.695 \,.\nn
\end{eqnarray}
It is observed that the first two orders are substantially smaller than in
eq.~\eqn{rnlb0MSb}, due to the fact that the no-pole term has effectively been
resummed into the global prefactor. The third order is of a similar size and
the 4th order turns out to be negative, which indicates that the leading
UV renormalon singularity is already dominating. This is confirmed by the
separated contributions of the lowest-lying IR and UV renormalons, again
provided in table~\ref{tab2}. This time large cancellations between the
lowest-lying UV and IR renormalons take place for the second and 4th order.
This cancellation could be the reason for an anomalously small second order
coefficient. Like in the $\MSb$ scheme, dominance of the leading UV renormalon
at $u=-1$ sets in at about the 5th order.

To conclude our discussion of the perturbative expansion of $\Psi^{''}(Q^2)$
in the large-$\beta_0$ approximation, we investigate the scheme with $C_m=C_a=0$
in both no-pole and renormalon contributions. The corresponding first few
perturbative coefficients read
\begin{eqnarray}
\label{rnlb0Ce0}
r_1^{\beta_0}(\Cz,\Cz) &=& 2 \,, \quad
r_2^{\beta_0}(\Cz,\Cz) \,=\, \Big(\,\frac{11}{12} - 2\zeta_3 \Big)
\beta_1 \,=\, -\,6.6935 \,, \nn \\
\mvs
r_3^{\beta_0}(\Cz,\Cz) &=& \Big(\,\frac{5}{6} + 2\zeta_3 \Big)
\beta_1^2 \,=\, 65.558 \,, \\
\mvs
r_4^{\beta_0}(\Cz,\Cz) \!&=&\! \Big(\,\frac{37}{32} -
\frac{15}{2}\zeta_5 \Big) \beta_1^3 \,=\, -\,603.31 \,. \nn
\end{eqnarray}
In this case, the leading UV renormalon dominates already from the lowest
order which is reflected in the sign-alternating behaviour of the perturbative
coefficients. Also the strong growth of the coefficients that signals the
asymptotic behaviour of the series is observed. As an amusing aside, we
remark that in this scheme, at each order $n>1$, only the highest possible
$\zeta$-function coefficients $\zeta(2\,[n/2+1]-1)$ arise, where $[x]$ denotes
the integer value of $x$. In table~\ref{tab3}, once again the contributions in
percent to the first 6 perturbative coefficients are presented. As indicated
above, in this scheme one finds that already the second coefficient
$r_2^{\beta_0}$ is largely dominated by the leading UV renormalon at $u=-1$,
and for still higher orders the series is fully dominated by this contribution.
The respective behaviour is also expected from the exponential factor
$\exp(-C_a u)$ in eq.~\eqn{BPsipp} which entails that in the scheme with
$C_a=0$ the residues of the IR renormalon poles are no longer enhanced with
respect to the UV ones as is the case in the $\MSb$ scheme.

\begin{table}[htb]
\begin{center}
\begin{tabular}{lrrrrrr}
\hline\hline
& $\quad r_1^{\beta_0}\;$ & $\quad r_2^{\beta_0}\;$ &
  $\quad r_3^{\beta_0}\;$ & $\quad r_4^{\beta_0}\;$ &
  $\quad r_5^{\beta_0}\;$ & $\quad r_6^{\beta_0}\;$ \\
\hline
${\rm UV}_{-1}$ & 66.7 & 134.5 & 103.0 & 105.7 & 101.5 & 101.3 \\
${\rm UV}_{-2}$ & -16.7 & -22.4 & -9.0 & -4.7 & -2.3 & -1.2 \\
${\rm IR}_{2}$ & 50.0 & -16.8 & 3.9 & -1.4 & 0.5 & -0.2 \\
${\rm IR}_{3}$ & -7.4 & -1.7 & 0.8 & -0.3 & 0.1 & 0.0 \\
${\rm IR}_{4}$ & 4.2 & 1.4 & -0.4 & 0.1 & 0.0 & 0.0 \\
\hline
SUM & 96.8 & 95.0 & 98.2 & 99.4 & 99.8 & 99.9 \\
\hline\hline
\end{tabular}
\caption{Contribution (in percent) of the lowest-lying ultraviolet (UV) and
infrared (IR) renormalon poles to the first 6 perturbative coefficients
$r_n^{\beta_0}$ in the scheme with $C_m=C_a=0$ for both quark mass and
renormalon terms.\label{tab3}}
\end{center}
\end{table}

In an analogous fashion to the derivation of eq.~\eqn{Psipplb0res}, we can
derive an expression for the correlation function $D^L(Q^2)$ of eq.~\eqn{DLres}
in the large-$\beta_0$ approximation, which reads
\begin{eqnarray}
\label{DLlb0res}
D_{\beta_0}^L(Q^2) &=& -\,\frac{N_c}{8\pi^2}\,
\wh m^2 \,[\alpha_s^{C_m}(Q)]^{2\gamma_m^{(1)}/\beta_1} \biggl\{\, 1 -
2\,\frac{\gamma_m^{(1)}}{\beta_1} \ln\Big[ 1 + C_m \sfrac{\beta_1}{2}\,
a_Q^{C_m} \Big] \nn \\
\mvs
&& +\,\frac{2\pi}{\beta_1}\! \int\limits_0^\infty \!{\rm d}u\,
{\rm e}^{-2u/(\beta_1 a_Q^{C_a})}\cdot \frac{3C_F}{2\pi}\,{\rm e}^{-C_a u}\,
G_D(u) + \ldots \,\biggr\} \,.
\end{eqnarray}
The perturbative expansion of this correlator shall only be discussed in the
mixed scheme with $C_m=0$ for the quark mass and $\MSb$, that is $C_a=-5/3$,
for the remainder. Then, the coefficients $\tilde r_n$ of eq.~\eqn{DLmhat} in
the large-$\beta_0$ limit are found as
\begin{eqnarray}
\label{rntlb0}
\tilde r_1^{\beta_0}(\Cz,\MSb) &=& 4 \,, \quad
\tilde r_2^{\beta_0}(\Cz,\MSb) \,=\, \Big(\,\frac{25}{4} -
2\zeta_3\Big) \beta_1 \,=\, 17.3065 \,, \nn \\
\mvs
\tilde r_3^{\beta_0}(\Cz,\MSb) &=& \Big(\,\frac{205}{18} -
\frac{10}{3}\zeta_3 \Big) \beta_1^2 \,=\, 149.486 \,, \\
\mvs
\tilde r_4^{\beta_0}(\Cz,\MSb) &=& \Big(\,\frac{21209}{864} -
\frac{25}{6}\zeta_3 - \frac{15}{2}\zeta_5 \Big) \beta_1^3 \,=\, 1071.81 \,, \nn
\end{eqnarray}
where like before the numerical values have been given at $N_f=3$. It is again
observed that the coefficients $\tilde r_n^{\beta_0}$ are substantially worse
behaved than the coefficients $r_n^{\beta_0}$.

\begin{table}[t]
\begin{center}
\begin{tabular}{lrrrrrr}
\hline\hline
& $\quad \tilde r_1^{\beta_0}\;$ & $\quad \tilde r_2^{\beta_0}\;$ &
  $\quad \tilde r_3^{\beta_0}\;$ & $\quad \tilde r_4^{\beta_0}\;$ &
  $\quad \tilde r_5^{\beta_0}\;$ & $\quad \tilde r_6^{\beta_0}\;$\\
\hline
${\rm UV}_{-1}$ & 33.3 & -5.8 & 9.5 & -8.6 & 8.2 & -10.6 \\
${\rm UV}_{-2}$ & -8.3 & -2.9 & -1.1 & -0.3 & -0.1 & 0.0 \\
${\rm UV}_{1}$ & 100.0 & 138.7 & 109.9 & 123.1 & 99.1 & 115.3 \\
${\rm IR}_{2}$ & -25.0 & -28.2 & -16.7 & -12.8 & -6.4 & -4.2 \\
${\rm IR}_{3}$ & -3.7 & -4.5 & -2.8 & -2.3 & -1.1 & -0.7 \\
\hline
SUM & 96.3 & 97.3 & 98.8 & 99.2 & 99.7 & 99.8 \\
\hline\hline
& $\quad \tilde r_7^{\beta_0}\;$ & $\quad \tilde r_8^{\beta_0}\;$ &
  $\quad \tilde r_9^{\beta_0}\;$ & $\quad \tilde r_{10}^{\beta_0}\;$ &
  $\quad \tilde r_{11}^{\beta_0}\;$ & $\quad \tilde r_{12}^{\beta_0}\;$\\
\hline
${\rm UV}_{-1}$ & 9.6 & -13.2 & 11.3 & -16.2 & 13.1 & -19.4 \\
${\rm UV}_{-2}$ & 0.0 & 0.0 & 0.0 & 0.0 & 0.0 & 0.0 \\
${\rm UV}_{1}$ & 92.5 & 114.5 & 89.2 & 116.5 & 87.0 & 119.5 \\
${\rm IR}_{2}$ & -1.8 & -1.2 & -0.5 & -0.3 & -0.1 & -0.1 \\
${\rm IR}_{3}$ & -0.3 & -0.2 & -0.1 & 0.0 & 0.0 & 0.0 \\
\hline
SUM & 99.9 & 100.0 & 100.0 & 100.0 & 100.0 & 100.0 \\
\hline\hline
\end{tabular}
\caption{Contribution (in percent) of the lowest-lying ultraviolet (UV) and
infrared (IR) renormalon poles to the first 12 perturbative coefficients
$r_n^{\beta_0}$ in the mixed scheme with $C_m=0$ for the quark mass and
$\MSb$ in the renormalon terms.\label{tab4}}
\end{center}
\end{table}

Similarly to table~\ref{tab2}, in table~\ref{tab4} the contributions in
percent of the three lowest-lying UV renormalon poles at $u=1,-1,-2$ and two
lowest-lying IR renormalon poles at $u=2,3$ to the first 12 perturbative
coefficients $\tilde r_n$ in the large-$\beta_0$ approximation and the mixed
scheme are presented. The surprising finding that can also be inferred directly
from eq.~\eqn{DLlb0res} is that the function $D^L(Q^2)$ suffers from an
additional, spurious renormalon pole at $u=1$. This observation was, of course,
already made in ref.~\cite{bkm00}. Because the linear $u=1$ pole has the larger
residue as compared to the UV renormalon pole at $u=-1$, it dominates the
perturbative coefficients for a large number of orders, before the quadratic UV
pole at $u=-1$ takes over.\footnote{In the scheme with $C_m=C_a=0$, in which
the spurious pole at $u=1$ is less enhanced, still for many orders large
cancellations between the lowest-lying poles at $u=-1$ and $u=1$ take place.}

The origin of the renormalon pole at $u=1$ can be understood from eq.~\eqn{DL}.
In the construction of $D^L(Q^2)$, the term $\Psi(0)/s$ is subtracted. As will
be explained in more detail in appendix~\ref{appC}, the subtraction constant
$\Psi(0)$ consists of a contribution from the quark condensate and an UV
divergent perturbative term proportional to $m^4$. The subtraction of this
divergent term leads to an ambiguity which results in the emergence of the
additional renormalon at $u=1$, and since it is of UV origin, in
table~\ref{tab4} we have labelled the pole accordingly. Generally, in
applications, because of this spurious renormalon pole, it appears advisable
to avoid the correlator $D^L(Q^2)$ in phenomenological analyses.

A detailed discussion of the third physical observable related to the scalar
correlator, $\IM\Psi(s)$, in the large-$\beta_0$ limit, has been presented in
ref.~\cite{bkm00}, and therefore, we shall not repeat it here. We only remark
that, like $\Psi^{''}(s)$, also the spectral function does not suffer from a
renormalon pole at $u=1$. In the case of $\Psi^{''}(s)$, this pole contribution,
which is present in the independent perturbative coefficients $d_{n,1}$, is
cancelled by the term $2d_{n,2}$ (see eq.~\eqn{Psippres}), which individually
also receives contributions from a pole at $u=1$. In the case of $\IM\Psi(s)$,
those $u=1$ pole contributions are cancelled by the $(i\pi)^{2l}$ terms
multiplying $d_{n,2l+1}$ coefficients with $l\geq 1$ (see eq.~\eqn{ImPsi}).

To conclude, from the investigation of the scalar correlator in the
large-$\beta_0$ approximation, it appears advantageous to express at least the
global prefactor proportional to $\alpha_s^{2\gamma_m^{(1)}/\beta_1}$ in terms
of a scheme-invariant coupling $\hat\alpha_s$, such that the quark mass factor
is fully scheme independent. In the next section, we shall investigate the
options for such a definition of $\hat\alpha_s$ in full QCD and will study its
implications in section~\ref{sect5}.

\section{Scheme variations of the QCD coupling}\label{sect4}

The aim of this section is to define a class of renormalisation schemes in
which the running of the QCD coupling is scheme invariant, in particular it
only depends on the two leading $\beta$-function coefficients $\beta_1$ and
$\beta_2$. In addition, scheme transformations of this coupling can be
parametrised by just one parameter $C$, corresponding to transformations of
the QCD $\Lambda$-parameter, which sets the scale. Our starting point for the
construction of this class of couplings is the scale-invariant parameter
$\Lambda$ that can be defined as
\begin{equation}
\label{Lambda}
\Lambda \,\equiv\, Q\, {\rm e}^{-\frac{1}{\beta_1 a_Q}}
\,[ a_Q ]^{-\frac{\beta_2}{\beta_1^2}}
\exp\Biggl\{\,\int\limits_0^{a_Q}\,\frac{{\rm d}a}{\tilde\beta(a)}\Biggr\}\,,
\end{equation}
where
\begin{equation}
\frac{1}{\tilde\beta(a)} \,\equiv\, \frac{1}{\beta(a)} - \frac{1}{\beta_1 a^2}
+ \frac{\beta_2}{\beta_1^2 a} \,,
\end{equation}
which is free of singularities in the limit $a\to 0$. Consider a scheme
transformation to a new coupling $a'$, which takes the general form
\begin{equation}
\label{ap}
a' \,\equiv\, a + c_1\,a^2 + c_2\,a^3 + c_3\,a^4 + \ldots \,.
\end{equation}
The $\Lambda$-parameter in the new scheme, $\Lambda'$, only depends on $c_1$
and not on the remaining higher-order coefficients \cite{cg79}. The exact
relation between the $\Lambda$-parameters is given by
\begin{equation}
\label{Lambdap}
\Lambda' \,=\, \Lambda\,{\rm e}^{c_1/\beta_1} \,.
\end{equation}

This motivates the definition of a new coupling $\tilde a_Q$, which is
scheme invariant, except for shifts in the $\Lambda$-parameter, parametrised
by the constant $C$:
\begin{equation}
\label{atilde}
\frac{1}{\beta_1\tilde a_Q} \,\equiv\, \ln\frac{Q}{\Lambda} + \frac{C}{2} \,=\,
\frac{1}{\beta_1 a_Q} + \frac{C}{2} + \frac{\beta_2}{\beta_1^2} \ln a_Q -
\int\limits_0^{a_Q}\,\frac{{\rm d}a}{\tilde\beta(a)} \,.
\end{equation}
Like in the last section, we might have termed the new coupling $\tilde a_Q^C$,
in order to indicate its scheme dependence, but for notational simplicity,
we drop the superscript. In large-$\beta_0$ and the $\MSb$ scheme, the value
$C=-5/3$ led to the invariant construction of eq.~\eqn{ahatlb0}. As shall be
discussed further below, in full QCD the construction of a universal
scheme-invariant coupling appears not to be possible. The combination
\eqn{atilde} was already introduced in refs.~\cite{byz92,ben93}, where it was
noted that an unpleasant feature of $\tilde a_Q$ is the presence of the
non-analytic logarithmic term. However, we can get rid of it by an implicit
construction of another coupling $\hat a$, this time defined by
\begin{equation}
\label{ahat}
\frac{1}{\hat a_Q} \,\equiv\, \beta_1 \Big( \ln\frac{Q}{\Lambda} + \frac{C}{2}
\Big) - \frac{\beta_2}{\beta_1} \ln\hat a_Q \,=\,
\frac{1}{a_Q} + \frac{\beta_1}{2}\,C + \frac{\beta_2}{\beta_1}
\ln\frac{a_Q}{\hat a_Q} -
\beta_1 \!\int\limits_0^{a_Q}\,\frac{{\rm d}a}{\tilde\beta(a)} \,,
\end{equation}
which in perturbation theory should be interpreted in an iterative sense.

It is a straightforward matter to deduce from eq.~\eqn{ahat} the perturbative
relations that provide the transformations between the coupling $a$ in a
particular scheme and the coupling $\ah$.  Up to fourth order, taking $a$ as
well as the corresponding $\beta$-function coefficients in the $\MSb$ scheme,
and for $N_f=3$, we find
\begin{eqnarray}
\label{ahatofa}
\ah(a) &=& a - \sfrac{9}{4}\,C\,a^2 - \big( \sfrac{3397}{2592} + 4 C -
\sfrac{81}{16}\,C^2 \big) a^3 \nn \\
\tvs
&& -\,\big( \sfrac{741103}{186624} + \sfrac{233}{192}\,C - \sfrac{45}{2}\,C^2 +
\sfrac{729}{64}\,C^3 + \sfrac{445}{144}\zeta_3 \big) a^4 + \cO(a^5) \,,
\end{eqnarray}
as well as
\begin{eqnarray}
\label{aofahat}
a(\ah) &=& \ah + \sfrac{9}{4}\,C\,\ah^2 + \big( \sfrac{3397}{2592} + 4 C +
\sfrac{81}{16}\,C^2 \big) \ah^3 \nn \\
\tvs
&& +\,\big( \sfrac{741103}{186624} + \sfrac{18383}{1152}\,C + \sfrac{45}{2}\,C^2
+ \sfrac{729}{64}\,C^3 + \sfrac{445}{144}\zeta_3 \big) \ah^4 + \cO(\ah^5) \,.
\end{eqnarray}

As the next step, we investigate the running of the coupling $\ah$. To this
end, we first have to derive its $\beta$-function which is found to have the
simple form
\begin{equation}
\label{betahat}
-\,\mu\,\frac{{\rm d}\ah}{{\rm d}\mu} \,\equiv\, \hat\beta(\ah) \,=\,
\frac{\beta_1 \hat a^2}{\left(1 - \sfrac{\beta_2}{\beta_1}\, \hat a\right)} \,.
\end{equation}
Obviously, as is seen explicitly, it only depends on the scheme-invariant
$\beta$-function coefficients $\beta_1$ and $\beta_2$. However, our scheme is
different from the 't~Hooft scheme for which $\beta(a)=\beta_1 a^2+\beta_2 a^3$
\cite{tHo77}. We also note that non-trivial zeros of $\hat\beta(\ah)$ can only
arise in the case of $\beta_1=0$. Integrating the RGE \eqn{betahat}, one obtains
\begin{equation}
\label{ahatrun}
\frac{1}{\ah_Q} \,=\, \frac{1}{\ah_\mu} + \frac{\beta_1}{2}\ln\frac{Q^2}{\mu^2}
- \frac{\beta_2}{\beta_1}\ln\frac{\ah_Q}{\ah_\mu} \,.
\end{equation}
Again, this implicit equation for $\ah_Q$ can either be solved iteratively,
to provide a perturbative expansion, or, of course, numerically. In the
following section, we shall investigate the phenomenological implications of
re-expressing the perturbative expansion in terms of $\ah$ for the scalar
correlation function.

Before turning to the phenomenological applications, however, we point out
the possibility of defining a fully scheme-invariant coupling. Since the QCD
coupling is not directly measurable, such a definition would have to be based
on a particular physical observable, for example the QCD Adler function.
In the past, such definitions have been discussed in the literature. (See
e.g.~refs.~\cite{gru80,gru84}.) However, then the definition of the coupling is
non-universal and its $\Lambda$-parameter and $\beta$-function depend on the
perturbative expansion coefficients of the physical quantity. For this reason,
we prefer to stick to the universal coupling $\ah$ according to the definition
\eqn{ahat}, and study the behaviour of physical observables under variation of
the parameter $C$.

\section{Phenomenological applications}\label{sect5}

Let us now investigate the phenomenological implications of introducing the
QCD coupling $\wh\alpha_s$ of eq.~\eqn{ahat}. We begin by doing this on the
basis of the scalar correlator $\Psi_{\rm PT}^{''}$ of eq.~\eqn{Psippmhat},
where, as a first step, the coupling in the prefactor, originating in the
running of the quark mass, is re-expressed in terms of $\wh\alpha_s$. Defining
the quantity $\wh\Psi^{''}(\as)$, which just contains the dependence on the
coupling,
\begin{equation}
\label{PsippSI}
\Psi_{\rm PT}^{''}(Q^2) \,\equiv\, \frac{N_c}{8\pi^2}\,
\frac{\wh m^2}{Q^2}\; \wh\Psi^{''}(\as) \,,
\end{equation}
and employing the transformation of the QCD coupling provided in
eq.~\eqn{ahatofa}, we find:
\begin{eqnarray}
\label{psipphat}
\wh\Psi^{''}(\as) &=& [\wh\alpha_s(Q)]^{8/9} \big\{\, 1 + (5.4568 + 2\,C)\,a_Q
+ (25.452 + 14.469\,C - 0.25\,C^2)\,a_Q^2 \nn \\
\tvs
&+& (135.29 + 74.006\,C - 6.2531\,C^2 + 0.20833\,C^3)\,a_Q^3 \nn \\
\tvs
&+& (824.05 + 367.82\,C - 56.089\,C^2 + 9.2479\,C^3 - 0.24740\,C^4)
\,a_Q^4 + \ldots \big\} .
\end{eqnarray}
Thus far the coupling $a_Q$ within the curly brackets is left in the $\MSb$
scheme. We will proceed with investigating this case numerically and then, in
a second step, also rewrite these contributions in terms of $\ah_Q$.

\begin{figure}[thb]
\begin{center}
\includegraphics[height=6.4cm]{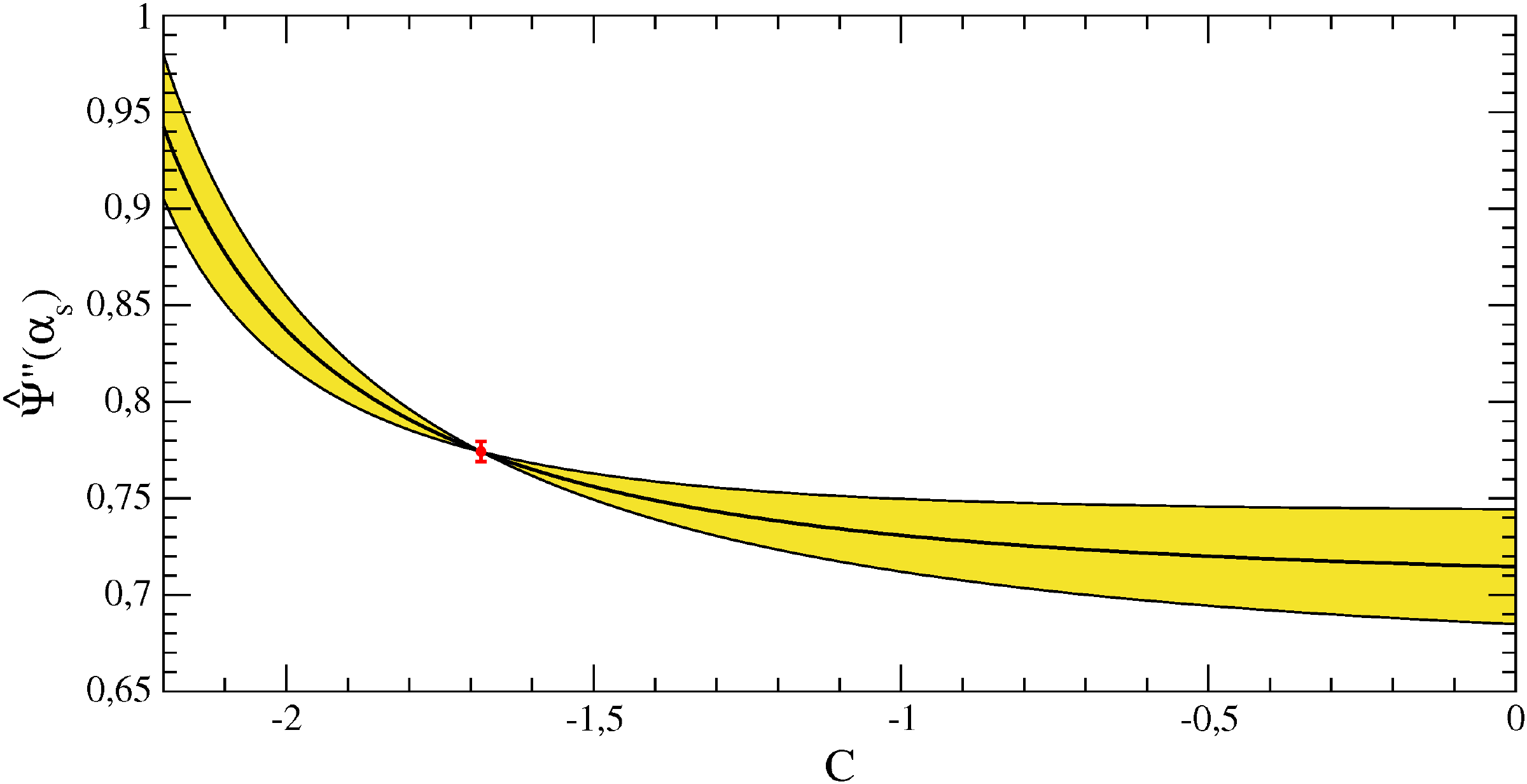}
\caption{$\wh\Psi^{''}(\as)$ according to eq.~\eqn{psipphat} as a function of
$C$ for $\as(M_\tau)=0.316$. The yellow band corresponds to either removing or
doubling the $\cO(a^4)$ correction to estimate the respective uncertainty. In
the red point, where $\cO(a^4)$ vanishes, the third order is taken as the error.
For further discussion, see the text.\label{fig1}}
\end{center}
\end{figure}

To this end, figure~\ref{fig1} displays a numerical account of the behaviour of
$\wh\Psi^{''}$ as a function of the scheme parameter $C$. As we are interested
in applications to hadronic $\tau$ decays in the future, for definiteness, we
have chosen $\as(M_\tau)=0.316$ in the $\MSb$ scheme, which corresponds to
the current PDG average $\as(M_Z)=0.1181$ \cite{pdg15}. The coupling
$\wh\alpha_s(Q)$ required in the prefactor has been determined by directly
solving eq.~\eqn{ahat} numerically, not via the expansion \eqn{ahatofa}.
In order to estimate the uncertainty in the perturbative prediction, the fourth
order term is either removed or doubled. The steepest curve in figure~\ref{fig1}
then corresponds to setting the $\cO(a_Q^4)$ contribution to zero and the
flattest one to doubling it. The yellow band hence corresponds to the region of
expected values for $\wh\Psi^{''}$, depending on the parameter $C$.

It is observed that at $C=-1.683$ the $\cO(a_Q^4)$ correction vanishes.
Interestingly enough, this value is surprisingly close to $C=-5/3$ in
large-$\beta_0$, which enters the construction of the invariant coupling
\eqn{ahatlb0} in the $\MSb$ scheme, though, presumably, this is merely a
coincidence. The red data point then indicates an estimate where the uncertainty
is taken to be the size of the third-order term. At this value of $C$, the
third-order correction has already turned negative and, beyond it, also the
$\cO(a_Q^4)$ contribution changes sign. This is an indication that in the
respective region of $C$ the contributions from IR and UV renormalons are more
balanced. To obtain a more complete picture, also the uncertainty of $\alpha_s$
should be folded in. From the PDG average $\as(M_Z)=0.1181(13)$ \cite{pdg15},
we deduce $\as(M_\tau)=0.316(10)$. Numerically, our result at $C=-1.683$ then
reads
\begin{equation}
\label{Oa4zero}
\wh\Psi^{''}(C=-1.683) \,=\, 0.774 \pm 0.005^{\,+\,0.058}_{\,-\,0.052}
\,=\, 0.774^{\,+\,0.058}_{\,-\,0.052} \,,
\end{equation}
where the first error corresponds to the $\cO(a_Q^3)$ correction also displayed
in figure~\ref{fig1}, while the second error results from the current
uncertainty in $\alpha_s$. The total error on the right-hand side has been
obtained by adding the individual uncertainties in quadrature.

The value \eqn{Oa4zero} can be compared to the result at $C=0$,
\begin{equation}
\label{Ceq0}
\wh\Psi^{''}(C=0) \,=\, 0.715 \pm 0.030^{\,+\,0.040}_{\,-\,0.038}
\,=\, 0.715^{\,+\,0.050}_{\,-\,0.048} \,.
\end{equation}
The two predictions \eqn{Oa4zero} and \eqn{Ceq0} are found to be compatible
and have similar uncertainties. At present, the error on $\alpha_s$ is dominant.
While in the prediction \eqn{Oa4zero}, the estimated uncertainty from missing
higher orders is substantially reduced, its sensitivity to $\alpha_s$ and
its uncertainty is increased. This is due to the fact that at $C=-1.683$,
symmetrising the error, one finds $\wh\alpha_s=0.610\pm 0.045$. This increased
sensitivity on $\alpha_s$ may also be seen as a virtue if one aims at an
extraction of $\alpha_s$ along the lines of \cite{bj08,bgjmmop12,bgmop14,pr16}.
In this respect, further understanding of the behaviour of the perturbative
series, for example, through models for the Borel transform in the spirit of
ref.~\cite{bj08}, could be helpful. As a last remark it is pointed out that
at the scale of $M_\tau$, for $C< -2$, the scheme transformation ceases to
be perturbative and breaks down. Therefore, such values should be discarded
for phenomenology.

\begin{figure}[thb]
\begin{center}
\includegraphics[height=6.4cm]{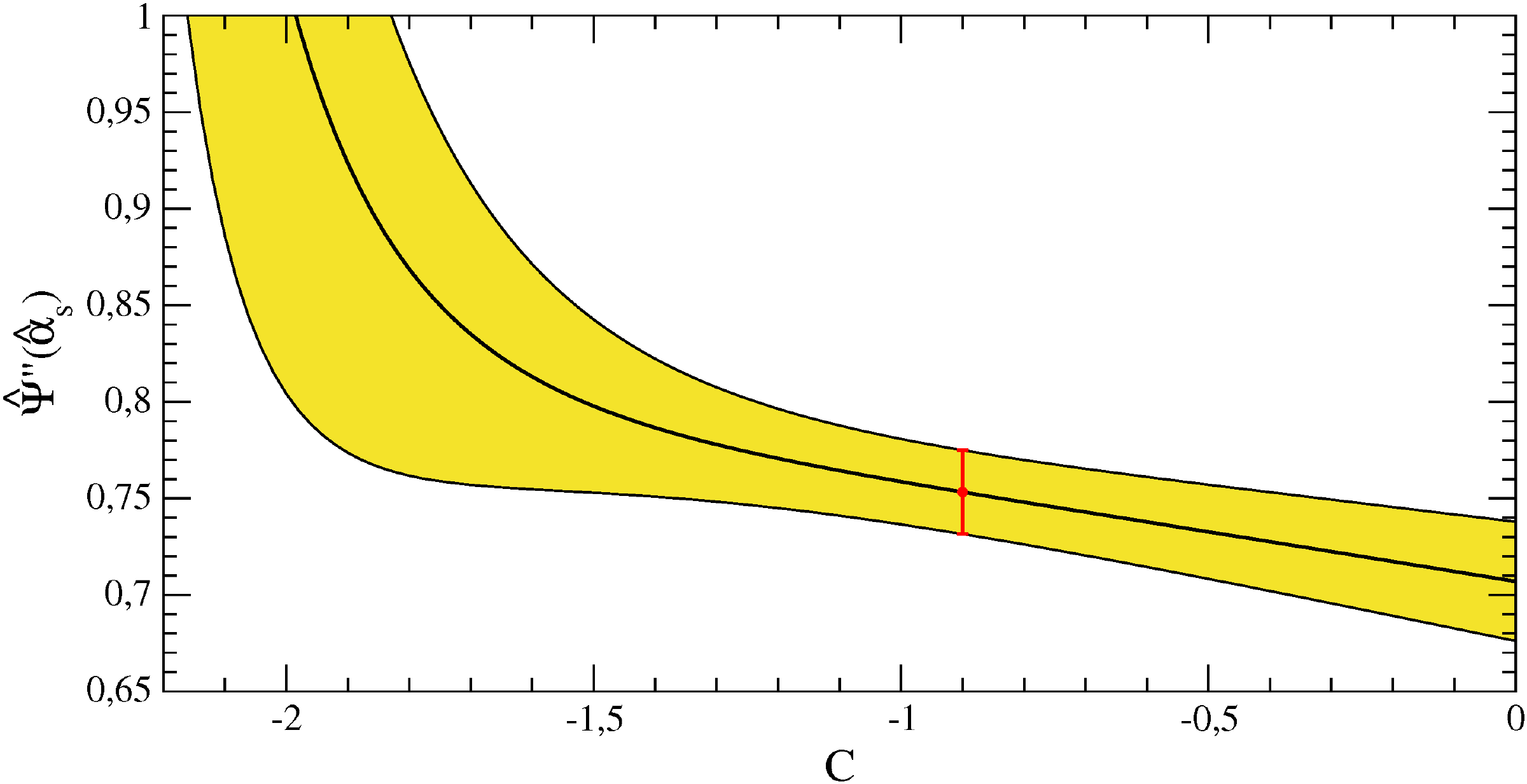}
\caption{$\wh\Psi^{''}(\wh\alpha_s)$ according to eq.~\eqn{psipphat2} as a
function of $C$ for $\as(M_\tau)=0.316$. The yellow band corresponds to either
removing or doubling the $\cO(\ah^4)$ correction to estimate the respective
uncertainty. At the red point, the uncertainty resulting from the $\cO(\ah^4)$
contribution is minimal. For further discussion, see the text.\label{fig2}}
\end{center}
\end{figure}

We proceed with our second step of also expressing the coupling $a_Q$ within
the curly brackets of eq.~\eqn{psipphat} in terms of $\ah_Q$. As a matter of
principle, we could introduce two different scheme constants $C_m$ and $C_a$,
related to mass and coupling renormalisation, respectively, since the global
prefactor originates from the quark mass, and the remaining expansion concerns
the QCD coupling. To keep the discussion more transparent, however, we prefer
to only use a single common constant $C=C_m=C_a$. Then the expansion in $\ah_Q$
takes the form  
\begin{eqnarray}
\label{psipphat2}
\wh\Psi^{''}(\wh\alpha_s) &=& [\wh\alpha_s(Q)]^{8/9} \big\{\, 1 + (5.4568 +
2\,C)\,\ah_Q + (25.452 + 26.747\,C + 4.25\,C^2)\,\ah_Q^2 \nn \\
\tvs
&+& (142.44 + 212.99\,C + 94.483\,C^2 + 9.2083\,C^3)\,\ah_Q^3 \nn \\
\tvs
&+& (932.71 + 1625.0\,C + 1099.8\,C^2 + 291.95\,C^3 + 20.143\,C^4)
\,\ah_Q^4 + \ldots \big\} .
\end{eqnarray}
The corresponding graphical representation of this result is displayed in
figure~\ref{fig2}. In this case, the order $\ah^4$ correction does not vanish
for any sensible value of $C$. The smallest uncertainty is assumed around
$C\approx -0.9$, at which one deduces
\begin{equation}
\label{Cmin}
\wh\Psi^{''}(C=-0.9) \,=\, 0.753 \pm 0.022^{\,+\,0.050}_{\,-\,0.046}
\,=\, 0.753^{\,+\,0.055}_{\,-\,0.051} \,.
\end{equation}
In figure~\ref{fig2}, the first error is shown as the red data point and
the second again corresponds to the uncertainty induced from the error on
$\alpha_s$. In view of the large $\alpha_s$ error, the result \eqn{Cmin} is
again fully compatible with \eqn{Oa4zero} and \eqn{Ceq0}.

\begin{figure}[thb]
\begin{center}
\includegraphics[height=6.4cm]{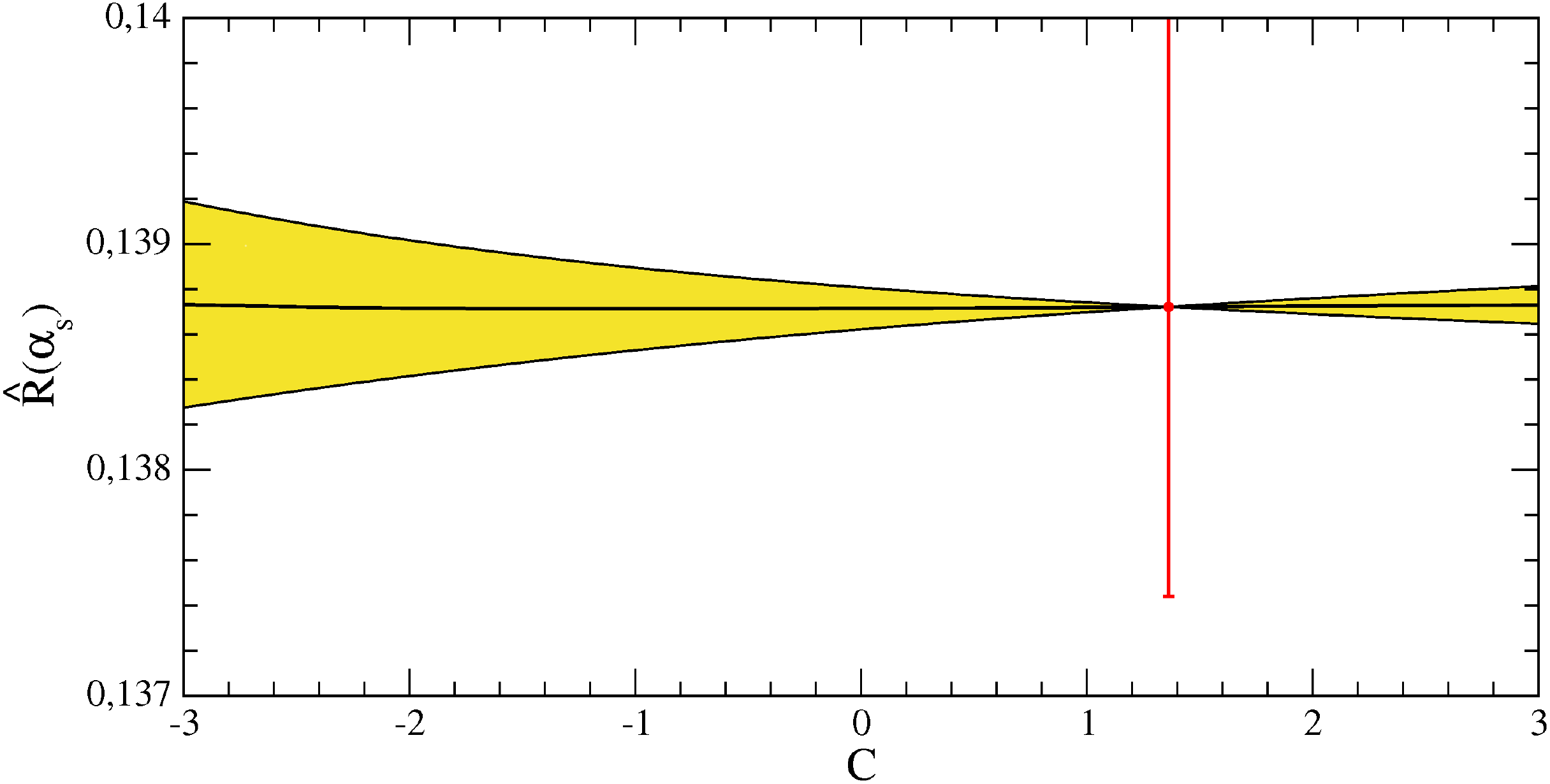}
\caption{$\wh R(\as)$ according to eq.~\eqn{Rhat} as a function of
$C$ for $\as(M_H)=0.1127$. The yellow band corresponds to either removing or
doubling the $\cO(a^4)$ correction to estimate the respective uncertainty. In
the red point, where $\cO(a^4)$ vanishes, the third order is taken as the error.
For further discussion, see the text.\label{fig3}}
\end{center}
\end{figure}

Let us now turn to the decay of the Higgs boson into quark-antiquark pairs.
The corresponding decay width is given by
\begin{equation}
\label{GammaH}
\Gamma(H\to q\bar q) \,=\, \frac{\sqrt{2}\,G_F}{M_H}\,
\IM\Psi\big(M_H^2+i0\big) \,\equiv\, \frac{N_c\,G_F M_H}{4\sqrt{2}\pi}\,
\wh m_q^2\, \wh R\big(\alpha_s(M_H)\big) \,,
\end{equation}
which defines the function $\wh R\big(\alpha_s(M_H)\big)$. We proceed in
analogy to the case of $\Psi^{''}$ by first expressing only the global
prefactor in terms of the coupling $\wh\alpha_s$, which results in
\begin{eqnarray}
\label{Rhat}
\wh R(\as) &=& [\wh\alpha_s(Q)]^{24/23} \big\{\, 1 + (8.0176 + 2\,C)\,a_Q
+ (46.732 + 18.557\,C + 0.08333\,C^2)\,a_Q^2 \nn \\
\tvs
&+& (142.12 + 117.09\,C - 1.5384\,C^2 - 0.05093\,C^3)\,a_Q^3 \nn \\
\tvs
&-& (544.67 - 426.17\,C + 22.522\,C^2 - 2.2856\,C^3 - 0.04774\,C^4)
\,a_Q^4 + \ldots \big\} .
\end{eqnarray}
Here the number of flavours $N_f=5$ and $Q=M_H$. For $\as(M_H)=0.1127$, a
graphical representation of $\wh R(\as)$ as a function of $C$ is given in 
figure~\ref{fig3}. Because the coupling now is much smaller than at the $\tau$
scale, the perturbative expansion converges faster, and thus the typical
$\cO(a^4)$ term is substantially smaller than the order $a^3$ correction at
$C=1.362$, where $\cO(a^4)$ vanishes. This is obvious from the large error bar
of the red point. The corresponding numerical result reads
\begin{equation}
\label{RhOa4zero}
\wh R(C=1.362) \,=\, 0.1387 \pm 0.0013 \pm 0.0020 \,=\, 0.1387 \pm 0.0024 \,,
\end{equation}
where the second error again results from the variation
$\alpha_s(M_H)=0.1127(12)$ which has been deduced from the PDG average. Still,
even though the large $\cO(a^3)$ uncertainty has been assumed, the current
error from the $\alpha_s$ input is even bigger.

\begin{figure}[thb]
\begin{center}
\includegraphics[height=6.4cm]{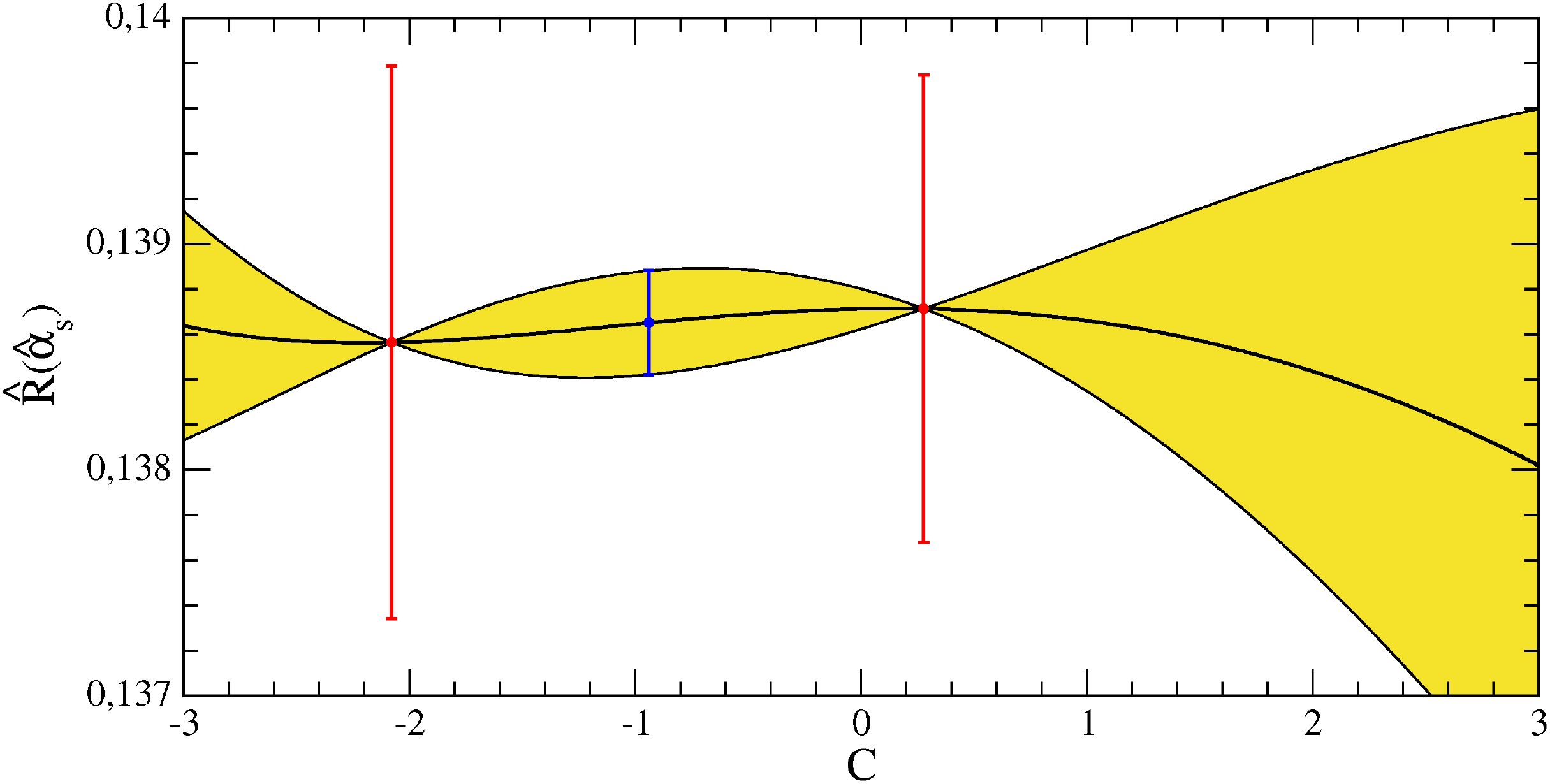}
\caption{$\wh R(\wh\alpha_s)$ according to eq.~\eqn{Rhat2} as a function of
$C$ for $\as(M_H)=0.1127$. The yellow band corresponds to either removing or
doubling the $\cO(\ah^4)$ correction to estimate the respective uncertainty.
In the red points, where $\cO(\ah^4)$ vanishes, the third order is taken as
the error. For further discussion, see the text.\label{fig4}}
\end{center}
\end{figure}

Like for $\Psi^{''}$, also for the Higgs decay, as a second step, we express
the remaining $\alpha_s$ series in powers of $\ah$. This yields
\begin{eqnarray}
\label{Rhat2}
\wh R(\wh\alpha_s) &=& [\wh\alpha_s(Q)]^{24/23} \big\{\, 1 + (8.0176 + 2\,C)
\,\ah_Q + (46.732 + 33.924\,C + 3.9167\,C^2)\,\ah_Q^2 \nn \\
\tvs
&+& (141.19 + 315.38\,C + 103.88\,C^2 + 7.6157\,C^3)\,\ah_Q^3 \nn \\
\tvs
&-& (524.03 - 1491.9\,C - 1353.1\,C^2 - 277.97\,C^3 - 14.756\,C^4)
\,\ah_Q^4 + \ldots \big\} ,
\end{eqnarray}
and the corresponding behaviour as a function of $C$ is presented in
figure~\ref{fig4}. This time two values of $C$ are found, at which the
$\cO(\ah^4)$ correction vanishes, and they are again displayed as the red data
points. In both cases, like before the corresponding uncertainty inferred from
the size of the third order is much larger than a typical fourth order term.
The corresponding numerical results are given by
\begin{equation}
\label{Rh2Oa4zero1}
\wh R(C=-2.079) \,=\, 0.1386 \pm 0.0012 \pm 0.0020 \,=\, 0.1386 \pm 0.0023 \,,
\end{equation}
and
\begin{equation}
\label{Rh2Oa4zero2}
\wh R(C=0.277) \,=\, 0.1387 \pm 0.0010 \pm 0.0020 \,=\, 0.1387 \pm 0.0022 \,,
\end{equation}
where the second error once more is due to the $\alpha_s$ uncertainty and the
final errors result from a quadratic average. In a situation like this, in our
opinion a conservative estimate of higher-order corrections can be obtained
by assuming the maximal $\cO(\ah^4)$ correction between those two points and
taking that as the perturbative uncertainty. This approach is shown as the blue
point, and the numerical value reads 
\begin{equation}
\label{Rh2Oa4max}
\wh R(C=-0.94) \,=\, 0.1387 \pm 0.0002 \pm 0.0020 \,=\, 0.1387 \pm 0.0020 \,.
\end{equation}
It is clear that now the higher-order uncertainty is completely negligible
with respect to the present error in $\alpha_s$.

To summarise, rewriting the perturbative expansion in terms of the coupling
$\wh\alpha_s$ of eq.~\eqn{ahat} introduces interesting approaches to improve
the convergence of the series for the known low-order corrections, before
the asymptotic behaviour sets in. We demonstrated this explicitly for the
correlator $\Psi^{''}(s)$ at the scale $M_\tau$ and for the decay of the Higgs
boson into quarks which is related to $\IM\Psi(s)$ at the scale $M_H$. In
both examples, however, the parametric uncertainty induced by the error on
$\alpha_s$ dominates. This is in part due to the recent increase in the
$\alpha_s$ uncertainty of the PDG average \cite{pdg15} by more than a factor
of two, in view of an earlier analysis of $\alpha_s$ determinations from
lattice QCD by the FLAG collaboration \cite{flag14}. Hence, we expect our
findings to increase in importance when the uncertainty on $\alpha_s$ again
shrinks in the future. Still, in view of the potential to strengthen the
sensitivity on $\alpha_s$, our approach could also open promising options for
improved non-lattice $\alpha_s$ determinations.

\section{Conclusions}\label{sect6}

The scalar correlation function is one of the basic QCD two-point correlators
with important phenomenological applications for the decay of the Higgs boson
to quark-antiquark pairs \cite{djou08}, determinations of light quark masses
from QCD sum rules \cite{jop02,jop06} and contributions to hadronic decays of
the $\tau$ lepton \cite{pp98,pp99,gjpps03}. Presently, the perturbative
expansion of the scalar correlator is known up to order $\alpha_s^4$ in the
strong coupling \cite{bck05}.

Three physical functions related to the scalar correlator play a role for
phenomenological studies: $\IM\Psi(s)$ in Higgs decay, $\Psi^{''}(s)$ for
quark-mass extractions and $D^L(s)$ in finite-energy sum rule analyses of
hadronic $\tau$ decays. From the known perturbative coefficients it is observed
that the renormalisation-group resummed $D^L(s)$ only depends on the independent
coefficients $d_{n,1}$, and those corrections turn out much larger than the
ones for $\Psi^{''}(s)$ and $\IM\Psi(s)$, for which combinations of the
$d_{n,1}$ and $d_{n,k}$ with $k>1$ appear. The latter coefficients are
calculable from the renormalisation group and only depend on lower-order
$d_{n,1}$, $\beta$-function coefficients, and those of the mass anomalous
dimension.

In order to understand this pattern of higher-order corrections better,
we reviewed the results for the scalar correlator in the large-$\beta_0$
approximation \cite{bkm00}, and derived compact expressions for the correlators
$\Psi^{''}(s)$ and $D^L(s)$ in terms of Borel transforms, which directly give
access to the renormalon structure of the respective correlators. While this
structure in the case of $\Psi^{''}(s)$ is analogous to the one of the Adler
function, double and single IR renormalon poles for $u\geq 2$, with only a
single pole at $u=2$, as well as double and single UV poles for $u\leq -1$,
for the correlator $D^L(s)$ an additional single pole at $u=1$ is found. The
origin of this spurious pole, which is suspected to be of UV origin, can be
traced back to the divergent subtraction $\Psi(0)/s$ that is performed in the
construction of $D^L(s)$. While the pole at $u=1$ is present in the coefficients
$d_{n,1}$, for $\Psi^{''}(s)$ and $\IM\Psi(s)$ it is cancelled by corresponding
contributions to the dependent coefficients $d_{n,k}$ with $k>1$.

Another feature of the scalar correlator that becomes apparent from the
large-$\beta_0$ approximation is the appearance of a regular contribution that 
is related to the renormalisation of the global mass factor $m^2$. By rewriting
this prefactor in terms of the renormalisation-group invariant quark mass
$\wh m$, one is left with the logarithmic term in eq.~\eqn{Psipplb0res}, which
depends on the leading-order RG coefficients $\beta_1$ and $\gamma_m^{(1)}$, as
well as the renormalisation scheme of the coupling in the prefactor. Expressing
this prefactor in terms of the coupling $\wh\alpha_s$ of eq.~\eqn{ahatlb0},
which can be considered an invariant coupling in large-$\beta_0$, the regular
logarithmic contribution is resummed. Improvements in the behaviour of the
perturbative series were also discussed in section~\ref{sect3}, and it was
concluded that this is in part due to shifting the contribution of UV
renormalon poles, in particular the lowest-lying one at $u=-1$, to lower orders.
Generally, however, it has to be acknowledged that for the scalar correlator
the large-$\beta_0$ limit does not provide a satisfactory representation of
the full QCD case.

In order to mimic as much as possible the large-$\beta_0$ case, in
section~\ref{sect4}, we attempted to define a scheme-invariant coupling also
for full QCD. Whereas it appears to be impossible to do this in a universal
way, that is, independent of any observable, in eq.~\eqn{ahat} we presented the
definition of a coupling $\wh\alpha_s$ whose running is renormalisation-group
invariant in the sense that it only depends on the invariant coefficients
$\beta_1$ and $\beta_2$, and is given by the simple $\beta$-function of
eq.~\eqn{betahat}. The scheme dependence of $\wh\alpha_s$ is then parametrised
by a single parameter $C$ which corresponds to transformations of the QCD
scale parameter $\Lambda$.

Phenomenological applications of re-expressing the perturbative series of 
$\Psi^{''}(s)$ at the $\tau$-mass scale, and $\IM\Psi(s)$ at $M_H$, in terms
of $\wh\alpha_s$, were investigated in section~\ref{sect5}. To this end, we
considered two cases: a first, in which only the $\alpha_s$-prefactor,
originating from the quark mass, is rewritten in $\wh\alpha_s$, and the
remaining series is kept in the $\MSb$ scheme, and a second case, in which the
whole series is expressed in terms of the coupling $\wh\alpha_s$. Generally,
it can be concluded that appropriate choices of $C$ allow for an improvement
of the behaviour of the perturbative series for the first few known orders.
This is, however, achieved at the expense of an increase in the value of the
coupling, either only in the prefactor, or also in the remaining expansion
terms, which leads to an increased sensitivity to $\alpha_s$ and also its
uncertainty.

In an era in which just recently the error on the PDG average of the strong
coupling \cite{pdg15} has increased by more than a factor of two, in view of
an earlier analysis of $\alpha_s$ determinations from lattice QCD by the FLAG
collaboration \cite{flag14}, we find that in all considered cases the
uncertainty of our perturbative predictions is dominated by the error on
$\alpha_s$. Therefore, in the investigated examples, currently, improvements
in the perturbative series appear to be a secondary issue. Still, when our
knowledge on the value of $\alpha_s$ at some point returns to a precision
comparable to previous estimates, the uncertainty due to higher-order
corrections becomes of a similar size, and optimising the series by
appropriate scheme choices through variation of the parameters $C$ should
allow for refined perturbative predictions.

On the other hand, the increased sensitivity on $\alpha_s$ for certain
ranges of $C$ can also be taken as a virtue if one aims at determinations of
$\alpha_s$, for example from hadronic $\tau$ decay spectra along the lines
of refs.~\cite{bj08,bgjmmop12,bgmop14,pr16}, as this could result in reduced
equivalent uncertainties in the $\MSb$ coupling. A preliminary assessment of
such an approach is performed in ref.~\cite{bjm16}, for the perturbative
expansion of the Adler function and the total $\tau$ hadronic width, before we
embark on a full-fledged analysis of the decay spectra. In this respect, also
analysing models for the Borel transform in the coupling $\wh\alpha_s$, along
the lines of ref.~\cite{bj08}, could provide additional helpful insights.

Since a substantial part of the improvements results from rewriting global
prefactors of $\alpha_s$, investigating other observables which include such
factors and suffer from large perturbative corrections could be rather
promising. These factors may either be explicitly present, like for example
in gluonium correlation functions which carry a global factor $\alpha_s^2$, or
may emerge from quark-mass factors, similarly to the scalar correlator, as in
the case of the total semi-leptonic $B$-meson decay rate which is proportional
to $m_b^5$. It is to be expected that also in these applications the
perturbative expansion could be improvable by adequate scheme choices for the
coupling $\wh\alpha_s$.

\bigskip
\acknowledgments
Helpful discussions with Martin~Beneke, Diogo~Boito and Antonio~Pineda are
gratefully acknowledged. This work has been supported in part by MINECO Grant
numbers CICYT-FEDER-FPA2011-25948 and CICYT-FEDER-FPA2014-55613-P, by the
Severo Ochoa excellence program of MINECO, Grant number SO-2012-0234, and
Secretaria d'Universitats i Recerca del Departament d'Economia i Coneixement
de la Generalitat de Catalunya under Grant number 2014 SGR 1450.

\appendix
\section{Renormalisation group functions and dependent coefficients}\label{appA}

In our notation, the QCD $\beta$-function and mass anomalous dimension are
defined as:
\begin{eqnarray}
\label{bega}
-\,\mu\,\frac{{\rm d}a}{{\rm d}\mu} &\equiv& \beta(a) \,=\,
\beta_1\,a^2 + \beta_2\,a^3 + \beta_3\,a^4 + \beta_4\,a^5 + \ldots \,, \\
\tvs
-\,\frac{\mu}{m}\,\frac{{\rm d}m}{{\rm d}\mu} &\equiv& \gamma_m(a) \,=\,
\gamma_m^{(1)}\,a + \gamma_m^{(2)}\,a^2 + \gamma_m^{(3)}\,a^3 +
\gamma_m^{(4)}\,a^4 + \ldots \,.
\end{eqnarray}
It is assumed that we work in a mass-independent renormalisation scheme
and in this study throughout the modified minimal subtraction scheme $\MSb$
is used. To make the presentation self-contained, below the known coefficients
of the $\beta$-function and mass anomalous dimension in the given conventions
shall be provided. Numerically, for $N_c=3$ the first four coefficients of the
$\beta$-function are given by \cite{tvz80,lrv97,cza04,bck16}
\begin{eqnarray}
\label{bfun}
\beta_1 &=& \sfrac{11}{2} - \sfrac{1}{3}\,N_f \,, \qquad
\beta_2 \,=\, \sfrac{51}{4} - \sfrac{19}{12}\,N_f \,, \qquad
\beta_3 \,=\, \sfrac{2857}{64} - \sfrac{5033}{576}\,N_f +
              \sfrac{325}{1728}\,N_f^2 \,, \nn \\
\tvs
\beta_4 &=& \sfrac{149753}{768} + \sfrac{891}{32}\,\zeta_3 -
\left(\sfrac{1078361}{20736} + \sfrac{1627}{864}\,\zeta_3\right) N_f +
\left(\sfrac{50065}{20736} + \sfrac{809}{1296}\,\zeta_3\right) N_f^2 +
\sfrac{1093}{93312}\,N_f^3 \,, \nn \\
\tvs
\beta_5 &=& \sfrac{8157455}{8192} + \sfrac{621885}{1024}\,\zeta_3 -
\sfrac{88209}{1024}\,\zeta_4 - \sfrac{144045}{256}\,\zeta_5 \nn \\
\tvs
&-& \left(\sfrac{336460813}{995328} + \sfrac{1202791}{10368}\,\zeta_3 -
\sfrac{33935}{3072}\,\zeta_4 - \sfrac{1358995}{13824}\,\zeta_5 \right) N_f \nn\\
\tvs
&+& \left(\sfrac{25960913}{995328} + \sfrac{698531}{41472}\,\zeta_3 -
\sfrac{5263}{2304}\,\zeta_4 - \sfrac{5965}{648}\,\zeta_5 \right) N_f^2 \nn \\
\tvs
&-& \left(\sfrac{630559}{2985984} + \sfrac{24361}{62208}\,\zeta_3 -
\sfrac{809}{6912}\,\zeta_4 - \sfrac{115}{1152}\,\zeta_5 \right) N_f^3 +
\left(\sfrac{1205}{1492992} - \sfrac{19}{5184}\,\zeta_3 \right)\,N_f^4 \,,
\end{eqnarray}
and the first five for $\gamma_m(a)$ are found to be \cite{vlr97,bck14}
\begin{eqnarray}
\label{gfun}
\gamma_m^{(1)} &=& 2 \,, \qquad
\gamma_m^{(2)} \,=\, \sfrac{101}{12} - \sfrac{5}{18}\,N_f \,, \qquad
\gamma_m^{(3)} \,=\, \sfrac{1249}{32} - \left(\sfrac{277}{108} +
                     \sfrac{5}{3}\,\zeta_3\right) N_f -
                     \sfrac{35}{648}\,N_f^2
\,, \nn \\
\tvs
\gamma_m^{(4)} &=& \sfrac{4603055}{20736} + \sfrac{1060}{27}\,\zeta_3 -
\sfrac{275}{4}\,\zeta_5 - \left(\sfrac{91723}{3456} + \sfrac{2137}{72}\,
\zeta_3 - \sfrac{55}{8}\,\zeta_4 - \sfrac{575}{36}\,\zeta_5 \right) N_f \nn \\
\svs
&+& \left(\sfrac{2621}{15552} + \sfrac{25}{36}\,\zeta_3 - \sfrac{5}{12}\,
\zeta_4 \right) N_f^2 - \left(\sfrac{83}{7776} -
\sfrac{1}{54}\,\zeta_3 \right) N_f^3 \,. \nn \\
\tvs
\gamma_m^{(5)} &=& \sfrac{99512327}{82944} + \sfrac{23201233}{62208}\,\zeta_3 +
\sfrac{3025}{16}\,\zeta_3^2 - \sfrac{349063}{2304}\,\zeta_4 -
\sfrac{28969645}{15552}\,\zeta_5 + \sfrac{15125}{32}\,\zeta_6 +
\sfrac{25795}{32}\,\zeta_7 \nn \\
\svs
&-& \left(\sfrac{150736283}{746496} + \sfrac{391813}{1296}\,\zeta_3 +
\sfrac{2365}{144}\,\zeta_3^2 - \sfrac{1019371}{6912}\,\zeta_4 -
\sfrac{12469045}{31104}\,\zeta_5 + \sfrac{39875}{288}\,\zeta_6 +
\sfrac{56875}{432}\,\zeta_7 \right) N_f \nn \\
\svs
&+& \left(\sfrac{660371}{186624} + \sfrac{251353}{15552}\,\zeta_3 +
\sfrac{725}{216}\,\zeta_3^2 - \sfrac{41575}{3456}\,\zeta_4 -
\sfrac{33005}{5184}\,\zeta_5 + \sfrac{2875}{432}\,\zeta_6 \right) N_f^2 \nn \\
\svs
&+& \left(\sfrac{91865}{746496} + \sfrac{803}{2592}\,\zeta_3 +
\sfrac{7}{72}\,\zeta_4 - \sfrac{10}{27}\,\zeta_5 \right) N_f^3 -
\left(\sfrac{65}{31104} + \sfrac{5}{1944}\,\zeta_3 - \sfrac{1}{216}\,\zeta_4
\right) N_f^4 \,.
\end{eqnarray}

The dependent perturbative coefficients $d_{n,k}$ with $k>1$ can be expressed
in terms of the independent coefficients $d_{n,1}$, and coefficients of the
QCD $\beta$-function and mass anomalous dimension. In particular, the
coefficients $d_{n,2}$, which are required in eq.~\eqn{Psippres}, take the form
\begin{equation}
\label{dn2}
d_{n,2} \,=\, -\,\frac{1}{2}\,\gamma_m^{(n)} d_{0,1} - \frac{1}{4}
\sum\limits_{k=1}^{n-1} \big( 2\gamma_m^{(n-k)} + k\,\beta_{n-k} \big)
d_{k,1} \,.
\end{equation}
Explicitly, at $N_c=3$ and up to the fourth order, they read:
\begin{eqnarray}
\label{d12tod42}
d_{1,2} &=& -\,1 \,, \qquad
d_{2,2} \,=\, -\,\sfrac{53}{3} + \sfrac{11}{18}\,N_f \,, \nn \\
\tvs
d_{3,2} &=& -\,\sfrac{49349}{144} + \sfrac{585}{8} \zeta_3 +
\left( \sfrac{11651}{432} - \sfrac{59}{12} \zeta_3 \right) N_f -
\left( \sfrac{275}{648} - \sfrac{1}{9} \zeta_3 \right) N_f^2 \,, \nn \\
\tvs
d_{4,2} &=& -\,\sfrac{49573615}{6912} + \sfrac{535759}{192} \zeta_3 -
\sfrac{30115}{96} \zeta_5 + \left( \sfrac{56935973}{62208} - 
\sfrac{243511}{864} \zeta_3 + \sfrac{5}{6} \zeta_4 + \sfrac{1115}{48} \zeta_5
\right) N_f \nn \\
&& -\, \left( \sfrac{6209245}{186624} - \sfrac{250}{27} \zeta_3 +
\sfrac{25}{36} \zeta_5 \right) N_f^2 + \left( \sfrac{985}{2916} -
\sfrac{5}{54} \zeta_3 \right) N_f^3 \,.
\end{eqnarray}

\section{The coefficients \boldmath{$\cD_n^{(1)}$} and \boldmath{$H_n^{(1)}$}}\label{appB}

Here, we provide the coefficients $\cD_n^{(1)}$ and $H_n^{(1)}$, required
to predict the perturbative coefficients $d_{n,1}$ in the large-$\beta_0$
approximation up to fifth order.

\begin{eqnarray}
\cD_1^{(1)}  &=&  -\,1 \,, \quad
\cD_2^{(1)} \,=\, -\,\sfrac{22}{3} \,, \quad
\cD_3^{(1)} \,=\, -\,\sfrac{275}{6} + 12\,\zeta_3 \,, \quad
\cD_4^{(1)} \,=\, -\,\sfrac{7880}{27} + 80\,\zeta_3 \,, \nn \\
\tvs
\cD_5^{(1)}  &=&  -\,\sfrac{324385}{162} + \sfrac{1000}{3}\,\zeta_3 +
600\,\zeta_5 \,, \quad
\cD_6^{(1)} \,=\, -\,\sfrac{1224355}{81} + \sfrac{10000}{9}\,\zeta_3 +
6000\,\zeta_5 \,.
\end{eqnarray}

\begin{eqnarray}
H_2^{(1)}  &=&  \sfrac{51}{2} \,, \quad
H_3^{(1)} \,=\, -\,\sfrac{585}{8} + 18\,\zeta_3 \,, \quad
H_4^{(1)} \,=\, \sfrac{15511}{72} - 54\,\zeta_3 \,, \nn \\
\tvs
H_5^{(1)}  &=&  -\,\sfrac{520771}{576} + \sfrac{585}{4}\,\zeta_3 
+ \sfrac{27}{4}\,\zeta_4 + 270\,\zeta_5 \,, \nn \\
\tvs
H_6^{(1)}  &=&  \sfrac{19577503}{4320} - \sfrac{2021}{6}\,\zeta_3 
- \sfrac{9}{2}\,\zeta_4 - \sfrac{8946}{5}\,\zeta_5 \,.
\end{eqnarray}

\section{The subtraction constant \boldmath{$\Psi(0)$}}\label{appC}

In order to understand the structure of the subtraction constant $\Psi(0)$,
examining the lowest perturbative order is sufficient. For definiteness,
we consider the case of the current \eqn{jtau} that plays a role in hadronic
$\tau$ decays. $\Psi(0)$ receives contributions from the normal-ordered quark
condensate and a perturbative term proportional to $m^4$. At lowest order it
reads:
\begin{eqnarray}
\label{Psi0}
\Psi(0) &=& -\,(m_u - m_s)\big[\,
\langle\Omega|\!:\!\bar u u\!:\!|\Omega\rangle -
\langle\Omega|\!:\!\bar s s\!:\!|\Omega\rangle \,\big] \nn \\
\tvs
&& +\, 4iN_c\, (m_u - m_s) \big[\, m_u I_{m_u} - m_s I_{m_s} \,\big] \,,
\end{eqnarray}
where $I_m$ is the UV divergent massive scalar vacuum-bubble integral
\begin{equation}
I_m \,\equiv\, \mu^{2\ve}\!\!\int\!\frac{{\rm d}^D\!k}{(2\pi)^D}\,
\frac{1}{(k^2-m^2+i0)} \,=\, \frac{i}{(4\pi)^2}\,m^2\,\biggl\{\,
\frac{1}{\hat\ve} - \ln\frac{m^2}{\mu^2} + 1 + \cO(\ve) \,\biggr\} \,.
\end{equation}
The explicit expression for $I_m$ has been provided in dimensional
regularisation with $D=4-2\ve$ and $1/\hat\ve\equiv 1/\ve-\gamma_E+\ln(4\pi)$,
but the particular regularisation scheme is inessential for our argument.

Precisely the same massive scalar vacuum-bubble contribution as in the
second line of eq.~\eqn{Psi0} also arises when rewriting the normal-ordered
condensates in terms of non-normal-ordered minimally subtracted quark
condensates \cite{sc88,jm93}. Therefore, $\Psi(0)$ can also be expressed as
\begin{equation}
\label{Psi0nno}
\Psi(0) \,=\, -\,(m_u - m_s)\big[\,
\langle\Omega|\bar u u|\Omega\rangle -
\langle\Omega|\bar s s|\Omega\rangle \,\big] \,,
\end{equation}
which absorbs the mass logarithms in the definition of the quark condensate.
Due to a Ward identity, the condensate contribution in $\Psi(0)$ does not
receive higher-order corrections, and at least at next-to-leading order, it
has been checked that the perturbative term matches the vacuum-bubble structure
that arises when rewriting $\langle\Omega|\!:\!\bar qq\!:\!|\Omega\rangle$
in terms of $\langle\Omega|\bar qq|\Omega\rangle$ \cite{gen90}. It is expected
that this behaviour, and hence also the form of eq.~\eqn{Psi0nno}, should
remain the same to all orders. As an aside, it may be remarked that for the
pseudoscalar channel the combination \eqn{Psi0nno} with flavour sums of quark
masses as well as condensates is precisely what appears in the
Gell-Mann-Oakes-Renner relation \cite{gmor68,mj02}.

As we have seen, the subtraction constant $\Psi(0)$ suffers from a UV
divergence originating from the perturbative quark-mass correction in
eq.~\eqn{Psi0}. Even though this contribution can be absorbed in the definition
of the quark condensate by rewriting normal-ordered in terms of
non-normal-ordered condensates, because of the subtraction of $\Psi(0)/s$,
the UV divergence reflects itself in the spurious renormalon at $u=1$ in the
correlation function $D^L(Q^2)$ of eq.~\eqn{DLlb0res}.

\bigskip
\providecommand{\href}[2]{#2}\begingroup\raggedright\endgroup

\end{document}